\documentclass[11pt,letterpaper]{article}
\pdfoutput=1
\usepackage{geometry}
\usepackage{graphicx}
\usepackage{amssymb}
\usepackage{amsmath}
\usepackage{epstopdf}
\usepackage{mathabx}

\newcommand{\C}{{\mathbb C}}

\title{\vspace{1in} \bf\LARGE Instantons on Gravitons}
\author{\Large Sergey A. Cherkis\\
\\
\it School of Mathematics and Hamilton Mathematics Institute,\\
\it Trinity College Dublin,\  Dublin 2,\ Ireland\\
\\
\it Center for Theoretical Physics, Department of Physics,\\ 
\it University of California, Berkeley, CA 94720, USA\\
\\
\it Department of Mathematics, Stanford University,\\ 
\it Stanford, CA 94305, USA\\
\\
\it  \tt cherkis@maths.tcd.ie}
                                         
\begin{document}

\begin{titlepage}

\renewcommand{\thepage}{ }
\date{}

\maketitle

\abstract{Yang-Mills instantons on ALE gravitational instantons were constructed by Kronheimer and Nakajima in terms of matrices satisfying algebraic equations.  These were conveniently organized into a quiver.  
We construct generic Yang-Mills instantons on ALF gravitational instantons.  Our data are formulated in terms of matrix-valued functions of a single variable, that are in turn organized into a bow.  We introduce the general notion of a bow, its representation, its associated data and moduli space of solutions.  For a judiciously chosen bow the Nahm transform maps any bow solution to an instanton on an ALF space.  We demonstrate that this map respects all complex structures on the moduli spaces, so it is likely to be an isometry, and use this fact to study the asymptotics of the moduli spaces of instantons on ALF spaces.}
\vspace{-7in}

\noindent\parbox{\linewidth
}
{
\shortstack{
{\em\Large Dedicated to the memory of }
\hspace{4cm}   TCDMATH 10-04\\
{\em\Large
Israel Moiseevich Gelfand}
\hfill IHES/P/10/19\\
{\em\Large \phantom{A}} \hfill HMI 10-02}}

\end{titlepage}

\tableofcontents

\section{Introduction}
In the paper entitled ``Polygons and Gravitons'' \cite{Hitchin:1900zr} Hitchin rederives the Gibbons-Hawking form \cite{Hawking:1976jb, Gibbons:1979zt} of the hyperk\"ahler metrics on deformations of ${\mathbb R}^4/{\mathbb Z}_{k+1}$ from the corresponding twistor spaces.  These metrics are $A_k$ ALE spaces.  Similar techniques are used in \cite{Cherkis:1998hi} and \cite{Cherkis:2003wk} to obtain all ALF metrics.  In this paper we construct Yang-Mills instantons\footnote{An instanton is a finite action hermitian connection with self-dual curvature two-form.}  on these ALF spaces.  To this end we introduce bow diagrams, which are of interest in their own right.

The study of instantons involves a diverse number of techniques ranging from differential geometry and  integrability  to representation theory and string theory.  Since the construction by Kronheimer and Nakajima \cite{KN} of instantons on ALE spaces, the moduli spaces of these instantons emerged in a number of other areas of mathematics and theoretical physics. In \cite{KN} an instanton configuration is encoded in terms of quiver data, so that the instanton moduli spaces can be interpreted as quiver varieties.  The connections between instantons and the representation theory are particularly intriguing.  The relation between representations of Kac-Moody algebras and cohomology groups of instanton moduli spaces was discovered by Nakajima  \cite{Nakajima1994, Nakajima:1995ka}.  In \cite{Lusztig} quantum groups are constructed in terms of quivers.  More recently a version of the geometric Langlands duality for complex surfaces was formulated in \cite{BF, Licata, Nakajima08}, relating moduli spaces of instantons on ALE spaces, representations of affine Kac-Moody algebras, and double affine Grassmanians.  Such relations appeared in the physics literature in \cite{Dijkgraaf:2007sw} and \cite{Tan:2008wp}. 
A beautiful string theory derivation of this correspondence appeared recently in \cite{Witten:2009at}.  In fact, from the string theory picture of \cite{Witten:2009at} it is more natural to consider instantons on ALF, rather than on ALE, spaces.  The construction of such instantons is exactly the problem that we pursue in this paper.  We introduce the notion of a bow, thereby generalizing the notion of a quiver.  Just as for a quiver  we also introduce a representation of a bow and the moduli space of its representation.  These moduli spaces are richer and have larger $L^{2}$ cohomology.  We expect that these also have a representation-theoretic interpretation.

We would like to emphasize that almost any meaningful question about quiver varieties can be studied for the bow varieties we define here.  Any bow has a zero interval length limit in which it becomes a quiver.  Any representation of the limiting quiver appears as a limit of some representation of the original bow.  Not all bow representations, however,  become quiver representations in this limit.  As formulated in \cite{Cherkis:2009jm}, there is a certain reciprocity acting on the bows and their respective representations.  In particular, the representations that have a good quiver limit are exactly those for which the dual bow representation is balanced, as defined in Section~\ref{Sec:Balanced}.  For these bow representations the corresponding bow variety is isomorphic to the corresponding quiver variety\footnote{We ought to emphasize that as Riemannian manifolds these bow and quiver moduli spaces differ.}.  For all other representations, however, bow varieties appear to differ from quiver varieties.

\subsection{Background}

\subsubsection{General Constructions of Instantons}
All instanton configurations on flat ${\mathbb R}^{4}$ were constructed in  \cite{Atiyah:1978ri} by a technique now called the ADHM construction.  The ADHM data is encoded in terms of a quiver with one vertex and one edge.

This construction was generalized by Nahm in \cite{NahmCalorons}  to produce all instantons on ${\mathbb R}^{3}\times S^{1}.$  The Nahm data can no longer be interpreted as quiver data.  It finds a natural interpretation in terms of a particular simple bow.

Considering ADHM construction data invariant under the action of a finite subgroup $\Gamma$ of $SU(2)\subset SO(4),$  Kronheimer and Nakajima   \cite{KN} constructed instantons on deformations of ${\mathbb R}^{4}/\Gamma,$ which are called ALE gravitational instantons.

All of these constructions were rediscovered within string theory. See  \cite{Witten:1994tz}, \cite{Douglas} for the ADHM construction,  \cite{Douglas:1996sw} for the Kronheimer-Nakajima construction on $A_{k}$ ALE space,  \cite{Johnson:1996py} the Kronheimer-Nakajima construction on a general ALE space, and \cite{Diaconescu:1996rk}  for the Nahm construction.

A number of interesting  explicit instanton solutions based on these general constructions were obtained.  Some such examples are three instantons on ${\mathbb R}^4$ \cite{Korepin:1984bn},  instantons on Eguchi-Hanson space \cite{Bianchi:1995xd}, and  single caloron \cite{Kraan:1998xe, Lee:1998bb}, to name a few.

\subsubsection{Instantons on the Taub-NUT and on the multi-Taub-NUT}
Some instantons on the Taub-NUT space and even on multi-Taub-NUT space which have trivial holonomy at infinity are found in  \cite{Etesi:2001fb} and  \cite{Etesi:2002cc}.  It is proved in \cite{Etesi:2008ew} that this construction provides all instantons of instanton number one. This construction, unfortunately, is limited, since it is hard to introduce nontrivial monodromy at infinity and since it is not clear how to generalize it to generic configurations with larger instanton numbers.  It is very useful indeed in generating examples of large instanton number configurations.

A general construction of instantons on the Taub-NUT and on the multi-Taub-NUT spaces can be formulated in terms of $A$-type bows as in \cite{Cherkis:2008ip} and  \cite{Cherkis:2009jm}.  In particular, as an illustration of this construction, the  moduli space of one instanton on the Taub-NUT space is found in \cite{Cherkis:2008ip} and the explicit one instanton connection is found in  \cite{Cherkis:2009jm}.

In this work we aim to formulate a general construction for generic 
instantons on ALF spaces.  While an instanton on an ALE space is determined by matrices satisfying algebraic equations, an instanton on an ALF space is determined by matrix-valued functions satisfying ordinary differential equations.  If for an ALE space the corresponding ADHM data is conveniently organized into a quiver, for each ALF space we introduce a corresponding bow to serve the same purpose.

A quiver is a collection of points and oriented edges connecting some of them.  Given a quiver one can construct a corresponding bow.   In order to obtain this bow, consider a collection of oriented and parameterized intervals, each interval $I_{\sigma}$ corresponding to a point $\sigma $ of the original quiver.  If any two points $\sigma $ and $\rho$ of the quiver are connected by an edge oriented from $\sigma $ to $\rho$, we connect the corresponding intervals by an edge, so that this edge connects the right end of the former interval $I_{\sigma}$ to the left end of the latter interval $I_{\rho}$.  Sending the lengths of all of the intervals to zero reduces this bow to the original quiver.  

For a quiver there is a notion of a representation assigning a pair of vector spaces to each vertex \cite{Nakajima1994}.  Here we define a notion of a representation of a bow, which generalizes a quiver representation. Bow representations are richer:  if we consider the zero length limit in which a bow degenerates into a quiver, only some of the bow's representations  produce a limiting  conventional representations of a quiver.  This provides one of the motivations behind this work, as the new representations, as well as the moduli spaces associated with them, should carry additional information about self-dual Yang-Mills configurations and, one might expect, about representations of Kac-Moody algebras.

\subsection{Yang-Mills Instanton}
A connection $(d+A\wedge\ )$ on a Hermitian vector bundle over a Riemannian four-manifold ${\cal M}$ is said to be a Yang-Mills instanton if its curvature two-form $F=dA+A\wedge A$ is self-dual
\footnote{Whether one studies instantons satisfying $*F=F$ or anti-instantons satisfying $*F=-F$ is a matter of taste, as these two conditions are interchanged by a change of orientation of the base manifold.  To avoid any potential ambiguity, we specify that what we call an instanton here has the curvature two-form of type $(1,1)$ in every complex structure of the hyperk\"ahler base.} \label{Page:SDorASD}
under the action of the Hodge star operation:
\begin{equation}
F=*F,
\end{equation}
and if the Chern number of the corresponding bundle is finite:
\begin{equation}
\int\limits_{\cal M} {\rm tr}\, \ F\wedge F<\infty.
\end{equation}

Here we limit our consideration to the base manifold (which is also referred to as the background) ${\cal M}$ being a self-dual gravitational instanton.  Moreover, we only let ${\cal M}$ to be an ALF space as defined below.

\subsection{Self-dual Gravitational Instantons}
A self-dual gravitational instanton is a Riemannian manifold $({\cal M}, g)$ with the self-dual Riemann curvature two-form valued in ${\rm End}(T{\cal M})$:
\begin{equation}
R=*R,
\end{equation}
and finite Pontrjagin number 
\begin{equation}
\int\limits_{\cal M} {\rm tr}\ R\wedge R<\infty.
\end{equation}
There are only two kinds of compact self-dual gravitational instantons: the flat four-torus $T^4$ and the $K3$ surface.  Noncompact gravitational instantons can be distinguished by their asymptotic volume growth.   Choosing a point $x\in{\cal M}$ we denote by $B_x(r)$ the volume of a ball of radius $r$ centered at $x.$  If $\cal M$ is a self-dual gravitational instanton and if there are some positive constants $A$ and $B$ such that the volume of a ball satisfies 
\begin{equation}
A r^\nu\leq {\rm Vol} \, B_x(r)<B r^\nu,
\end{equation}
for $r>1,$  then we call the space $\cal M$
\begin{itemize}
\item an ALE space if $\nu=4,$  
\item an ALF space if $3\leq\nu<4,$
\item an ALG space if $2\leq\nu<3,$
\item and an ALH space if $0<\nu<2.$
\end{itemize}

As follows from Theorem 3.25 of \cite{Minerbe}, under the presumption $\int {\rm tr}\ r R\wedge R<\infty,$ each ALF space has $\nu=3$ and asymptotically its  metric has a local triholomorphic isometry.  

We conjecture that any ALF space (i.e. a manifold with self-dual curvature form, finite Pontrjiagin number, and cubic volume growth) is either a multi-Taub-NUT space \cite{mTN} (also called $A_k$ ALF space) or a $D_k$ ALF space \cite{Cherkis:1998xca, Cherkis:2003wk}.  For $A_k$ ALF space $k\geq-1$ and for $D_k$ ALF space $k\geq0.$  Both $A_k$ and $D_k$ ALF spaces have degenerate limits in which they can be viewed as the Kleinian singularities with 
\begin{description}
\item[-] $A_k$ ALF space given by $xy=z^{k+1}$ and 
\item[-] $D_k$ ALF space given by $x^2-z y^2=z^{k-1}$ 
\end{description}
as complex surfaces in ${\mathbb C}^3.$  Each of these singular spaces admits an ALF metric with one singular point, and a general ALF space is a smooth hyperk\"ahler deformation of one of these.  For the low values of $k$ in particular, the $A_0$ ALF is the Taub-NUT space, the $A_{-1}$ ALF is ${\mathbb R}^3\times S^1,$  the $D_2$  ALF is a deformation of $({\mathbb R}^3\times S^1)/{\mathbb Z}_2,$  the $D_1$ ALF is the deformation of the double cover of the Atiyah-Hitchin space \cite{Atiyah:1988jp} studied in \cite{Dancer:1992km}, while the $D_0$ ALF is the Atiyah-Hitchin space itself.

The study of Yang-Mills instantons on $A_k$ and $D_k$ ALF spaces is the main goal of this work.  Our construction of such instantons is formulated in terms of bow diagrams.  We introduce the notion of a bow and  its representation in Section~\ref{Sec:BowAll}, and  the moduli space of a bow representation  in Section \ref{Sec:Moduli}.  
In Section \ref{Sec:ALF} we realize ALF spaces as moduli spaces of certain bow representations. This is essential for the formulation of the Nahm transform.  Bow data defining an instanton are given in Section \ref{Sec:InstBow}, where the representation ranks are related to the instanton charges.  We claim that up to gauge equivalence this data is in one-to-one correspondence with the instanton. This correspondence is provided by the Nahm transform leading to the Yang-Mills instanton on ALF spaces given in Section \ref{Sec:NahmTransform}.  Section \ref{Sec:Cohomology} contains the proof of the self-duality condition for the constructed connections.  This proof makes it transparent that the Nahm transform, mapping between the moduli space of a bow representation and the moduli space of instantons, is an isomorphism of complex varieties in any of their complex structures.  We describe the bow moduli spaces as finite hyperk\"ahler quotients in Section~\ref{Sec:HKR} and compute their asymptotics in  Section~\ref{Sec:Asymp}.

\section{Generalizing Quivers}\label{Sec:BowAll}
In this section we introduce the notion of a bow that is at the center of our construction.  It is a generalization of the notion of a quiver, such that a quiver is a degenerate case of a bow.  As we aim to demonstrate it has richer structures associated with it.

\subsection{Bows}\label{Sec:Bows}
\begin{figure}[htbp]
\begin{center}
\includegraphics[width=0.4\textwidth]{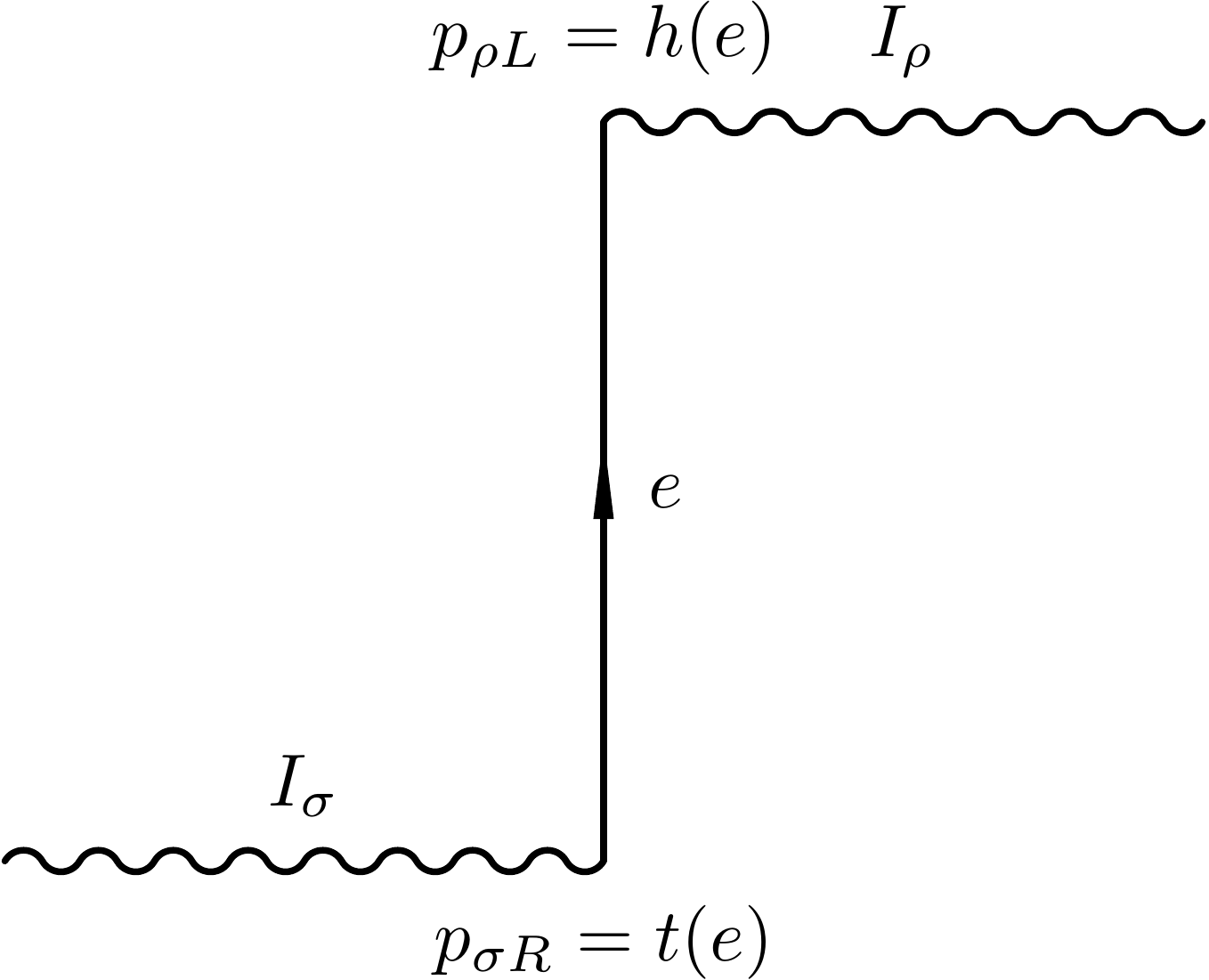}
\caption{An oriented edge $e$ connecting two intervals $I_\sigma$ and $I_\rho.$}
\label{Fig:Edge}
\end{center}
\end{figure}

A {\em bow} is defined by the following data.
\begin{itemize}
\item A collection of oriented closed distinct intervals ${\cal I}=\{I_\sigma\},$ for concreteness we let each interval be parameterized by $s$ with $p_{\sigma L}\leq s\leq p_{\sigma R}$ so that $I_\sigma=[p_{\sigma L}, p_{\sigma R}].$  We denote the length of each interval $I_\sigma$ by $l_\sigma,$ i.e. $l_\sigma=p_{\sigma R}-p_{\sigma L}.$
\item A collection ${\cal E}$ of oriented edges, such that each edge $e\in{\cal E}$ begins at the right end of some interval ${\cal I}_\sigma$ and ends at the left end of some, possibly the same, interval ${\cal I}_\rho$ as in Figure~\ref{Fig:Edge}.  Let $h(e)$ denote the head of the edge $e$ and $t(e)$ denote the tail, then 
\begin{equation}
A_{\sigma\rho}=\#\left\{e\in{\cal E} | t(e)=p_{\sigma R}, h(e)=p_{\rho L}\right\},
\end{equation}
is the number of edges originating at $I_\sigma$ and ending at $I_\rho.$
\end{itemize}
For any oriented edge $e\in{\cal E}$ let $\bar{e}$ denote this edge with the opposite orientation and $\bar{\cal E}=\{\bar{e}\}.$
When drawing a bow as a diagram, we use wavy lines to signify the intervals $I_\sigma$ and arrows to denote the edges.  A number of bow diagrams such as for example in Figs.~\ref{TNBow}, \ref{mTNBow}, and            \ref{DkALFBow} appear in Section  \ref{Sec:ALF} below.  It is clear from the definition that as all the lengths $l_\sigma\rightarrow 0$ a bow degenerates into a quiver.

\subsection{Representation of a Bow}
A regular {\em representation of a bow} is 
\begin{itemize}
\item A collection of distinct points $\Lambda=\{\lambda_\sigma^\alpha\}$ 
belonging to these intervals, such that $\lambda_\sigma^\alpha\in I_\sigma$ with $\alpha=1,2,\ldots, r_\sigma.$  A number $r_\sigma$ of $\lambda$-points belonging to an interval $I_\sigma$ can be zero.  
\item a collection of bundles $E\rightarrow{\cal I}$ consisting of Hermitian bundles
$$E_\sigma^0\rightarrow[p_{\sigma L}, \lambda_\sigma^1],\,  
E_\sigma^{r_\sigma}\rightarrow[\lambda_\sigma^{r_\sigma}, p_{\sigma R}],
{\rm and\ } 
 E_\sigma^\alpha\rightarrow[\lambda_\sigma^\alpha, \lambda_\sigma^{\alpha+1}]\  {\rm for\ } \alpha=1,2,\ldots,r_\sigma-1,$$
of ranks $R_\sigma^\beta={\rm rk}\, E_\sigma^\beta,$ satisfying the following matching conditions at $\lambda$-points:
\begin{align*}
 E_\sigma^\alpha|_{\lambda_\sigma^{\alpha+1}}\subset  E_\sigma^{\alpha+1}|_{\lambda_\sigma^{\alpha+1}}\ & {\rm if\ } R_\sigma^\alpha\leq R_\sigma^{\alpha+1}\ {\rm and\ } \\
  E_\sigma^\alpha|_{\lambda_\sigma^{\alpha+1}}\supset  E_\sigma^{\alpha+1}|_{\lambda_\sigma^{\alpha+1}}\ & {\rm if\ } R_\sigma^\alpha\geq R_\sigma^{\alpha+1}.
  \end{align*}
Let $\Lambda_0\subseteq\Lambda$ be the collection of $\lambda$-points at which $E$ does not change rank, i.e. $\Lambda_0=\{\lambda_\sigma^\alpha\in\Lambda\, |\, R_\sigma^{\alpha-1}= R_\sigma^{\alpha}\}.$
\item a collection of $\Lambda_0$-graded one-dimensional Hermitian spaces $W=\{W_\lambda\, |\, \lambda\in \Lambda_0 \}.$ 
 \end{itemize}

We call the number of $\lambda$-points $\#\Lambda=\sum_{\sigma} r_\sigma$ the {\em file of the representation,} and we call the collection $(R_\sigma^\alpha)$ the {\em rank of the representation.}

For simplicity we focus only on {\em regular} bow representations.  To define a general representation of a bow one can introduce multiplicities ${\cal W}=\{w_\sigma^\alpha\}$ of the $\lambda$-points with $w_\sigma^\alpha$ being the multiplicity of the points $\lambda_\sigma^\alpha$.  Denote by $w(\lambda)$ the multiplicity function, i.e. $w(\lambda_\sigma^\alpha)=w_\sigma^\alpha.$  In this case the assigned spaces in ${W}$ are such that for any $\lambda\in \Lambda_0$ the corresponding space dimension is given by the multiplicity: ${\rm dim}\, W_\lambda=w(\lambda).$ The matching conditions at $\Lambda$ however are more involved in this case and we postpone this discussion.  Here we focus only on regular bow representations.

\subsection{Bow Data}
In order to simplify our notation let $E_{\sigma R}\equiv E_\sigma^{r_\sigma}|_{p_{\sigma R}}$ denote the fiber over the right end of the interval $I_\sigma$ and $E_{\sigma L}\equiv E_\sigma^0|_{p_{\sigma L}}$ denote the fiber over the left end on the interval $I_\sigma,$ also, for some $\lambda=\lambda_\sigma^\alpha\in\Lambda_0$ let $E_\lambda\equiv  E_\sigma^{\alpha-1}|_{\lambda_\sigma^\alpha}=E_\sigma^{\alpha}|_{\lambda_\sigma^\alpha}$ denote the fiber at $\lambda.$  We denote the collections of the corresponding spaces by $E_R=\{E_{\sigma R}\}, E_L=\{E_{\sigma L}\},$ and $E_{\Lambda_0}=\{E_\lambda | \lambda\in\Lambda_0\}.$

To every bow representation  $\mathfrak R$ we associate a (generally infinite dimensional) hyperk\"ahler affine space ${\rm Dat}(\mathfrak{R})$.  It is a direct sum of three spaces ${\cal F}_{\rm in}\oplus{\cal F}_{\rm out},$\ ${\cal B}\oplus{\overline{\cal B}}$ and $\cal N$ that we call respectively fundamental, bifundamental, and Nahm spaces. 
We define the two components of the fundamental space by
\begin{align*}
{\cal F}_{\rm in}&={\rm Hom}(W,E_{\Lambda_0})\equiv\mathop{\oplus}_{\lambda\in\Lambda_0} {\rm Hom}(W_\lambda, E_\lambda),&
{\cal F}_{\rm out}&={\rm Hom}(E_{\Lambda_0},W)\equiv\mathop{\oplus}_{\lambda\in\Lambda_0} {\rm Hom}( E_\lambda,W_\lambda);
\end{align*}
the two components of the bifundamental space by
\begin{align*}
{\cal B}={\rm Hom}^{\cal E}(E_R,E_L)\equiv\mathop{\oplus}_{e\in{\cal E}} {\rm Hom}(E|_{t(e)}, E|_{h(e)}),\ 
\overline{\cal B}={\rm Hom}^{\overline{\cal E}}(E_L, E_R)\equiv\mathop{\oplus}_{e\in{\cal E}} {\rm Hom}(E|_{h(e)}, E|_{t(e)});
\end{align*}
and the Nahm space by
\begin{equation*}
{\cal N}={\rm Con}(E)\oplus{\rm End}(E)\otimes{\mathbb R}^3
\equiv\mathop{\oplus}_{\sigma, \alpha}\Big({\rm Con}(E_\sigma^\alpha)\oplus{\rm End}(E_\sigma^\alpha)\oplus{\rm End}(E_\sigma^\alpha)\oplus{\rm End}(E_\sigma^\alpha)\Big).
\end{equation*}
Here ${\rm Con}(E_\sigma^\alpha)$ is the space of connections $\nabla=\frac{d}{ds}-i T_0$ on the Hermitian bundle $E_\sigma^\alpha.$ We use the following notation to denote the {\em bow data}
\begin{equation}
\Big(I, J, B^{LR}, B^{RL},  (\nabla, T_1,T_2,T_3)\Big)\in
{\cal F}_{\rm in}\oplus{\cal F}_{\rm out}\oplus{\cal B}\oplus\overline{\cal B}\oplus{\cal N}={\rm Dat}(\mathfrak{R}).
\end{equation}

Now a comment is due specifying the behavior of the Nahm components at $\Lambda.$  These are the same conditions that appear in the Nahm transform of monopoles as formulated in \cite{Hurtubise:1990zf}.    At any $\lambda\in\Lambda_0$ the Nahm components have regular left and right limits at $\lambda.$ At $\lambda=\lambda_\sigma^\alpha\in\Lambda\setminus\Lambda_0$ on the other hand, with say $R_\sigma^{\alpha-1}<R_\sigma^\alpha,$ 
\footnote{For the case with $R_\sigma^{\alpha-1}>R_\sigma^\alpha$ the conditions are completely analogous.}
all $T_j(s)$ for $j=1,2,3$ have a regular limit from the left
\begin{equation}
\lim_{s\rightarrow \lambda-}T_j(s)=T_j^-(\lambda),
\end{equation}
while to the right of $\lambda$ we have
\begin{equation}\label{JumpMatch}
T_j(s)=\left(
\begin{array}{cc}
\frac{1}{2}\frac{\rho_j}{s-\lambda} +O((s-\lambda)^0) &O\Big((s-\lambda)^\frac{\Delta R-1}{2} \Big)\\
O\Big((s-\lambda)^\frac{\Delta R-1}{2} \Big)& T_j^-(\lambda)+O(s-\lambda)
\end{array}
\right),
\end{equation}
where $\Delta R=R_\sigma^\alpha-R_\sigma^{\alpha-1}$ is the change of rank and $(\rho_1, \rho_2, \rho_3)$ satisfy $[\rho_i, \rho_j]=2i\varepsilon_{ijk}\rho_k$ defining a $\Delta R$-dimensional irreducible representation of $su(2).$ 
\footnote{For the case of a general bow representation with a higher multiplicity point $\lambda_\sigma^\alpha,$ one of a number of possible conditions is that this representation splits into a direct sum of $w(\lambda_\sigma^\alpha)=w_\sigma^\alpha$ irreducible representations.}

\subsubsection{Gauge Group Action}
There is a natural gauge group action on the bow data.  We consider the group $\cal G$ of gauge transformations of $E:$ a gauge transformation $g\in{\cal G}$ is  smooth outside $\Lambda\setminus\Lambda_0$ and at $\lambda\in\Lambda\setminus\Lambda_0$ with $R_\sigma^{\alpha-1}<R_\sigma^\alpha$ satisfies
\begin{equation}
g(\lambda+)=
\left(\begin{array}{cc}
1 & 0\\
0 & g(\lambda-)
\end{array}\right).
\end{equation}
(The continuity condition for the case $R_\sigma^{\alpha-1}>R_\sigma^\alpha$ are analogous.)
The action of the gauge group on the bow data is
\begin{multline}\label{GaugeGroup}
g: \Big(I, J, B^{LR}, B^{RL},  (\nabla, T_1,T_2,T_3)\Big)
\mapsto\\
\Big(g^{-1}(\Lambda_0)I, Jg(\Lambda_0), g^{-1}_L B^{LR} g_R, g^{-1}_R B^{RL} g_L,  (g^{-1}\nabla g, g^{-1}T_1 g, g^{-1}T_2 g, g^{-1}T_3 g)\Big).
\end{multline}
Here for 
\begin{equation}
I=\mathop{\oplus}_{\lambda\in\Lambda_0} I_\lambda, \  
J=\mathop{\oplus}_{\lambda\in\Lambda_0} J_\lambda, \ 
B^{LR}=\mathop{\oplus}_{e\in{\cal E}} B^{LR}_e, \ 
B^{RL}=\mathop{\oplus}_{e\in{\cal E}} B^{RL}_e, 
\end{equation}
we use the natural conventions
\begin{align}
g^{-1}(\Lambda_0)I&=\mathop{\oplus}_{\lambda\in\Lambda_0} g^{-1}(\lambda) I_\lambda, &
g^{-1}_L B^{LR} g_R&=\mathop{\oplus}_{e\in{\cal E}} g^{-1}(h(e))B^{LR}_e g(t(e)), \\
Jg(\Lambda_0)&=\mathop{\oplus}_{\lambda\in\Lambda_0} J_\lambda g(\lambda), &
g^{-1}_R B^{RL} g_L&=\mathop{\oplus}_{e\in{\cal E}} g^{-1}(t(e))B^{RL}_e g(h(e)).
\end{align}
We shall also use the natural products of various collections such as for example 
\begin{align}
&IJ\equiv\mathop{\oplus}_{\lambda\in\Lambda_0} I_\lambda J_\lambda
\in {\rm End}\, E_{\Lambda_0} \equiv\mathop{\oplus}_{\lambda\in\Lambda_0} {\rm End}\, E_\lambda,\\
&B^{LR} B^{RL}\equiv\mathop{\oplus}_\sigma \sum_{\stackrel{e\in{\cal E}}{t(e)=p_{\sigma L}}} B^{LR}_e B^{RL}_e
\in{\rm End}\, E_L\equiv\mathop{\oplus}_\sigma {\rm End}\, E_{\sigma L},
\end{align}
and others.

\subsubsection{Hyperk\"ahler Structure}\label{Sec:HKStructure}
Each of the three spaces: the fundamental ${\cal F}_{\rm in}\oplus {\cal F}_{\rm out}$, the bifundamental ${\cal B}\oplus\overline{\cal B},$ and the Nahm ${\cal N},$ posesses a hyperk\"ahler structure specified below.  Let us now introduce a notation that makes this structure apparent.

Let $e_1, e_2, e_3$ be a two-dimensional representation of the quaternionic units, with $S$ being its representation space.  Thus, $e_j$ with $j=1,2,3$ satisfy the defining quaternionic relations $e_1^2=e_2^2=e_3^2=e_1 e_2 e_3=-1.$ For example one can choose a  representation in terms of the Pauli sigma matrices $e_j=-i\sigma_j;$ which, when written explicitly, is
\begin{equation}\label{Eq:Quaternions}
e_1=\left(\begin{array}{cc}
0 & -i\\
-i & 0
\end{array}\right),\ 
e_2=\left(\begin{array}{cc}
0 & -1\\
1 & 0
\end{array}\right),\  
e_3=\left(\begin{array}{cc}
-i & 0\\
0 & i
\end{array}\right). 
\end{equation}

Let us assemble the fundamental data into
\begin{align}
Q_\lambda&=\left(\begin{array}{c} J_\lambda^\dagger \\ I_\lambda   \end{array}\right): W_\lambda\rightarrow S\otimes E_\lambda&&{\rm and}& Q=\mathop{\oplus}_{\lambda\in\Lambda_0}Q_\lambda: W\rightarrow S\otimes E(\Lambda_0),
\end{align}
and the bifundamental data into 
\begin{align}
B_e^-&=\left(\begin{array}{c}  \big(B^{LR}_e\big)^\dagger\\ -B^{RL}_e  \end{array}\right): E_{t(e)}\rightarrow S\otimes E_{h(e)} &{\rm or\ }&&
B_e^+&=\left(\begin{array}{c}  \big(B^{RL}_e\big)^\dagger\\ B^{LR}_e  \end{array}\right): E_{h(e)}\rightarrow S\otimes E_{t(e)},
\end{align}
with 
\begin{align}
B^-&=\mathop{\oplus}_{e\in{\cal E}} B^-_e: E_L\rightarrow S\otimes E_R&&{\rm  and} &B^+&=\mathop{\oplus}_{e\in{\cal E}} B^+_e: E_R\rightarrow S\otimes E_L.
\end{align}
In order to simplify our notation and to avoid numerous brackets in our formulas we do not distinguish upper and lower $\pm$ indices, i.e. $B^+=B_+$ and $B^-=B_-.$ 
For the Nahm data $\nabla=\frac{d}{ds}-i T_0, T_1, T_2, T_3$ we introduce 
\begin{equation}
{\bf T}=1\otimes T_0+e_1\otimes T_1+e_2\otimes T_2+e_3\otimes T_3,
\end{equation}
and its quaternionic conjugate 
\begin{equation}
{\bf T}^*=1\otimes T_0-e_1\otimes T_1-e_2\otimes T_2-e_3\otimes T_3.
\end{equation}
If we understand $S\times I\rightarrow I$ to be a trivial bundle over the collection of intervals $I,$ then $\frac{d}{ds}-i{\bf T}$ is a connection on the bundle $S\otimes E.$

Now the quaternionic units act on the tangent space of ${\rm Dat}\,  {\mathfrak R}$ in this form by the left multiplication
\begin{equation}
e_j: (\delta Q, \delta B^-, \delta {\bf T})\mapsto 
\big((e_j\otimes 1_W)\delta Q, (e_j\otimes 1_{E_R}) \delta B^-, (e_j\otimes 1_{E})\delta {\bf T}\big).
\end{equation}
If one uses $B^+$ instead of $B^-$ to parameterize the bifundamental data the quaternionic unit action has the same form
\begin{equation}
e_j: (\delta Q, \delta B^+, \delta {\bf T})\mapsto 
\big((e_j\otimes 1_W)\delta Q, (e_j\otimes 1_{E_L}) \delta B^+, (e_j\otimes 1_{E})\delta {\bf T}\big).
\end{equation}

The space of bow data is a hyperk\"ahler space with the metric given by the direct product metric  
\begin{align}\label{metric}
ds^2&={\rm tr}_W\delta Q^\dagger\delta Q+{\rm tr}_{E_R}\delta B_+^\dagger\delta B_+
+\int \frac{1}{2}{\rm tr}_S {\rm tr}_E\delta{\bf T}^*\delta{\bf T} ds\nonumber\\
 &={\rm tr}_W\delta Q^\dagger\delta Q+{\rm tr}_{E_L}\delta B_-^\dagger\delta B_-
 +\int \frac{1}{2}{\rm tr}_S {\rm tr}_E\delta{\bf T}^*\delta{\bf T} ds.
\end{align}
As described above, the action of the three complex structures on $(\delta Q, \delta B_+,  \delta {\bf T})$ is by the left multiplication by $e_1, e_2,$ and $e_3.$  It leads to three corresponding K\"ahler forms $\omega_j(\cdot,\cdot)=g(\cdot, e_j\cdot),\ j=1,2,3,$ which can be organized into a purely imaginary quaternion 
\begin{equation*}
\omega= e_1\otimes \omega_1+e_2\otimes\omega_2+e_3\otimes \omega_3.
\end{equation*}
By direct computation
\begin{align}\label{Mom1}
\omega&={\rm Im}\,\left({\rm tr}_{E_{\Lambda_0}} \delta Q\wedge\delta Q^\dagger+{\rm tr}_{E_L}\delta B_+\wedge\delta B_+^\dagger
+\frac{1}{2} \int{\rm tr}_E\delta {\bf T}\wedge\delta {\bf T}^* ds\right)\\
\label{Mom2}
&={\rm Im}\,\left({\rm tr}_{E_{\Lambda_0}} \delta Q\wedge\delta Q^\dagger+{\rm tr}_{E_R}\delta B_-\wedge\delta B_-^\dagger
+\frac{1}{2} \int{\rm tr}_E\delta {\bf T}\wedge\delta {\bf T}^* ds\right).
\end{align} 

The metric \eqref{metric} is clearly compatible with the quaternionic structures and is invariant under the gauge group action \eqref{GaugeGroup}. The latter becomes apparent when we observe that \eqref{GaugeGroup} now takes the form 
\begin{equation}\label{Gauge}
g: \left(\begin{array}{c} 
Q   \\
B^+\\B^-\\ 
{\bf T}(s)\\ 
 \end{array}\right)
\mapsto
\left(\begin{array}{c} 
g^{-1}(\Lambda_0) Q\\
g^{-1}_{L}B^+g_{R}\\ g^{-1}_{R}B^- g_{L}\\ 
i g^{-1}(s)\frac{d}{ds}g(s)+g^{-1}(s){\bf T}(s)g(s)
    \end{array}\right).
\end{equation}

\subsection{Bow Solutions}\label{Sec:BowSolution}
Since the gauge group action of ${\cal G}$ preserves the hyperk\"ahler structure on the space of bow data, we can perform the hyperk\"ahler reduction \cite{Hitchin:1986ea}.  Namely, we can find the triplet of moment maps $(\mu_1, \mu_2, \mu_3),$ with each $\mu_j$ valued in the dual of the  Lie algebra, which we assemble into a pure imaginary quaternionic expression
\begin{equation*}
\mu=e_1\otimes\mu_1+e_2\otimes\mu_2+e_3\otimes\mu_3,
\end{equation*}
such that for any vector field $X$ generating an infinitesimal gauge transformation in ${\cal G}$ we have $d\mu(X)={\rm i}_X\omega,$ where ${\rm i}_X\omega$ is the interior product of the vector field $X$ and the two form $\omega$.  Using the definition and Eqs.~\eqref{Mom1} and \eqref{Mom2} one finds
\begin{multline}\label{MomentMaps}
\mu(Q, B, T)={\rm Im}\, (-i) \bigg(
i \frac{d}{ds}{\bf T}^*+{\bf T}{\bf T}^*
+\sum_{\lambda\in\Lambda_0}\delta(s-\lambda)  Q_\lambda Q^\dagger_\lambda\\
+\sum_{e\in{\cal E}}\left(\delta(s-t(e))  B^-_e \big(B^-_e\big)^\dagger
+\delta(s-h(e)) B^+_e \big(B^+_e\big)^\dagger\right)
\bigg).
\end{multline}

If we are to perform a hyperk\"ahler reduction, the value of the moment map $\mu(Q, B, T)$ has to be invariant under the gauge group action $\mu(s)\mapsto g^{-1}(s)\mu(s) g(s)$, i.e. it has to be an abelian character of the group of the gauge transformations.  It follows that we are to impose $\mu(Q,B,{\bf T})= e_1 \otimes  \nu_1(s) 1_E+e_2\otimes \nu_2(s) 1_E+e_3\otimes \nu_3(s) 1_E$ with each $\nu_1(s), \nu_2(s)$ and $\nu_3(s)$ some real-valued functions. Thus for any representation of a bow and a choice of a pure imaginary quaternion function $\nu=\nu_1 e_1+\nu_2 e_2+\nu_3 e_3$ we obtain the subset $\mu^{-1}(\nu)$ in the space of bow data which inherits the isometric action of the gauge group $\cal G$ on it.  The subset $\mu^{-1}(\nu)$ is called the {\em level set} at level $\nu.$  The space of gauge group orbits ${\cal M}(\nu)=\mu^{-1}(\nu)/{\cal G}={\rm Dat}(\mathfrak{R})/\!\!/\!\!/{\cal G}$ is the quotient hyperk\"ahler space.  We call this space the {\em moduli space} of the bow representation or simply the moduli space of the bow, if it is clear from the context which representation is being considered.  

A point of ${\cal M}(\nu)$ is a gauge equivalence class of some bow data $(Q, B, {\bf T})$ satisfying $\mu(Q, B, {\bf T})=\nu.$  We call such data a {\em bow solution} in representation $\mathfrak{R}$ at level $\nu.$ 

\subsection{Bow Cohomology}
Let us use the following isometry on the space of bow data to simplify our moment map values:
\begin{equation}\label{shift}
{\bf T}(s)\mapsto{\bf T}(s)-\int^s\nu(s')ds'.
\end{equation} 
If $(Q, B, {\bf T})$ was satisfying the moment map conditions \eqref{MomentMaps} with $\mu=\nu(s),$ after the above redefinition ${\bf T}_{\rm New}(s)={\bf T}(s)+\int^s\nu(s')ds'$ the data $(Q, B, {\bf T}_{\rm New})$ satisfies \eqref{MomentMaps} with $\mu=\sum\limits_\sigma \left(\delta(s-p_{\sigma R})\int\limits^{p_{\sigma R}}\nu(s')ds'-\delta(s-p_{\sigma L})\int\limits^{p_{\sigma L}}\nu(s')ds'\right).$  Thus it suffices to study the values of the moment maps of the form 
$\nu=\sum\limits_\sigma \left(\delta(s-p_{\sigma R})\nu_{\sigma L}-\delta(s-p_{\sigma L})\nu_{\sigma R}\right).$

As a matter of fact, we shall choose the value of $\mu$ to be a bow cocycle. Imposing the cocycle condition allows us to choose the value of $\mu$ to be 
\begin{equation}\label{Eq:CohForm}
\sum_{e\in{\cal E}}\big(\delta(s-t(e))-\delta(s-h(e))\big) \nu_e 1_E,
\end{equation}
so that the level is given by a pure quaternionic imaginary function $\nu_e$ on the set of edges. 
In order to do this let us outline what we mean by the bow cohomology.

Given a bow with the collections of intervals $I$ and edges $\cal E,$ let ${\cal C}(I)$ denote the space of smooth real functions on $I,$ let ${\cal C}({\cal E})={\mathbb R}^{\# \cal E}$ denote the space of functions on the set of edges, and let $L$ denote the set of closed paths.  A closed path is a cyclically ordered alternating  sequence of intervals and edges such that each edge connects two intervals that are adjacent to it in this sequence. We let ${\cal C}(L)$ denote the space of functions on the space of closed paths.

Now let us define the space of 0-cochains to be $C^0={\cal C}(I),$ the space of 1-cochains to be $C^1={\cal C}(I)\oplus {\cal C}({\cal E}),$ and the space of 2-cochains to be $C^2={\cal C}(L).$  For $f,g\in{\cal C}(I), a\in{\cal C}({\cal E}), $ and $r\in{\cal C}(L)$ we define the differentials in the complex
\begin{equation}\label{complex}
\Gamma: 0\rightarrow C^0\xrightarrow{d_0}C^1\xrightarrow[]{d_1}C^2\rightarrow0\end{equation}
by
\begin{align}
d_0: f(s)\mapsto
\left(\begin{array}{c}  g(s)\\ a(e)  \end{array}\right)=&
\left(\begin{array}{c}   \frac{d}{ds}f(s)\\ f(h(e))-f(t(e)) \end{array}\right),&& \\
 &d_1:\left(\begin{array}{c}  g(s)\\ a(e)  \end{array}\right)\mapsto
r(l)=\sum_{\sigma | I_\sigma\in l}\int_{I_\sigma} g(s)ds+\sum_{e\in l} a(e).
\end{align}

Since for an edge connecting $I_\sigma$ to $I_\rho$ we have  $t(e)=p_{\sigma R}$ and $h(e)=p_{\rho L}$ from our definition of a closed path it follows that $d_1 d_0=0$ and \eqref{complex} is a cochain complex of a bow. 0-cocycles are functions on a bow constant on each connected component.   In the context of our moment map discussion above the Nahm data redefinitions
\begin{equation}
T'_j(s)=T_j(s)+f_j(s),
\end{equation}
produces the change in the moment map
\begin{equation}
\mu(Q, B, {\bf T}')=\mu(Q, B, {\bf T})+d_0 \left(\sum_{j=1}^3 e_j f_j\right),
\end{equation}
that is a 1-coboundary.  So, any two levels that differ by a coboundary the moduli spaces at these levels are isometric.  We impose the 1-cocycle condition on our choice of the level to ensure the absence of noncommutativity of the resulting instanton base space.  Thus the  first cohomology of a bow $H^1(\Gamma)$ parameterizes essentially different moment map components and we can consider the values of $\mu\in H^1(\Gamma)\oplus H^1(\Gamma)\oplus H^1(\Gamma).$  Each such value can be chosen to be of the form \eqref{Eq:CohForm}.

\subsection{Various Forms of the Moment Map Conditions}
One can write the  moment map condition 
\begin{equation}\label{Eq:MoMa}
\mu(Q, B, {\bf T})= \sum_{e\in{\cal E}}\big(\delta(s-t(e))-\delta(s-h(e))\big) \nu_e,
\end{equation}
with the moment map given in Eq.~\eqref{MomentMaps} in a number of forms.  Depending on the context we can use one or another of these forms.  In the remainder of this section we write the moment map conditions \eqref{Eq:MoMa} in four different forms: the quaternionic, the complex, the real, and the operator form.

\subsubsection{Quaternionic Form}
In the interior of each subinterval
\begin{equation}
{\rm Im}\,\left(\frac{d}{ds}{\bf T}+i{\bf T}{\bf T}^*\right)=0,
\end{equation}
at a point $\lambda\in\Lambda_0$ we have 
\begin{equation}
{\rm Im}\,{\bf T}(\lambda+)-{\rm Im}\,{\bf T}(\lambda-)={\rm Im}\,(-i) Q_\lambda Q_\lambda^\dagger,
\end{equation}
while at $\lambda\in\Lambda\setminus\Lambda_0$ we have the conditions \eqref{JumpMatch}.

At the ends of each interval $I_\sigma$ we obtain the conditions
\begin{align}
{\rm Im}\,\, {\bf T}(p_{\sigma R})&=\sum_{\stackrel{e\in{\cal E}}{t(e)=p_{\sigma R}}}\left( {\rm Im}\, i\, B^-_e\big(B^-_e\big)^\dagger-\nu_e\right),\\
{\rm Im}\,\, {\bf T}(p_{\sigma L})&=\sum_{\stackrel{e\in{\cal E}}{h(e)=p_{\sigma L}}} \left( {\rm Im}\,(-i)  B^+_e\big(B^+_e\big)^\dagger-\nu_e\right).
\end{align}

\subsubsection{The Complex Form}\label{Sec:Clx}
Letting $D=\frac{d}{ds}-i T_0+T_3, T=T_1+i T_2$ and $\nu^{\mathbb C}=\nu_1+i\nu_2$ 
\begin{align}\label{ClxEq}
[D, T]&-\sum_{\lambda\in\Lambda_0}\delta(s-\lambda)I_\lambda J_\lambda\\
&+\sum_{e\in{\cal E}}\bigg(\delta(s-t(e))\Big(B^{RL}B^{LR}+\nu^{\mathbb C}\Big)-\delta(s-h(e))\Big(B^{LR}B^{RL}+\nu^{\mathbb C}\Big)\bigg)=0, \\
\nonumber
[D^\dagger, D]+[T^\dagger, T]&+\sum_{\lambda\in\Lambda_0}\delta(s-\lambda)(J_\lambda^\dagger J_\lambda-I_\lambda I_\lambda^\dagger)\\
& +\sum_{e\in{\cal E}}\Bigg(\delta(s-t(e))\left((B_e^{LR})^\dagger B_e^{LR}-B_e^{RL}( B_e^{RL})^\dagger-2\nu_{e3}\right)\\
 \label{RealEq}
&+\delta(s-h(e))\left((B_e^{RL})^\dagger B_e^{RL}-B^{LR} (B^{LR})^\dagger+2\nu_{e3}\right)\Bigg)=0.
\end{align}

\subsubsection{Real Form}
Inside each interval outside $\Lambda$ the Nahm data satisfies the Nahm equations
\begin{align}
\frac{d}{ds} T_1&=i[T_0, T_1]+i[T_2, T_3],\\
\frac{d}{ds} T_2&=i[T_0, T_2]+i[T_3, T_1],\\
\frac{d}{ds} T_3&=i[T_0, T_3]+i[T_1, T_2],
\end{align}
at $\lambda\in\Lambda_{0}$
\begin{align}
T_1(\lambda+)-T_1(\lambda-)&=\frac{1}{2}\left(I_\lambda J_\lambda+J_\lambda^\dagger I_\lambda^\dagger\right),\\
T_2(\lambda+)-T_2(\lambda-)&=\frac{i}{2}\left(J_\lambda^\dagger I_\lambda^\dagger-I_\lambda J_\lambda\right),\\
T_3(\lambda+)-T_3(\lambda-)&=\frac{1}{2}\left(J_\lambda^\dagger J_\lambda-I_\lambda I_\lambda^\dagger\right),
\end{align}
while at the ends of an interval $I_{\sigma}$
\begin{align}
T_1(p_{\sigma L})&=\sum_{\stackrel{e\in{\cal E}}{h(e)=p_{\sigma L}}}
\left(\frac{1}{2}\left(B_e^{LR} B_e^{RL}+(B_e^{RL})^\dagger (B_e^{LR})^\dagger\right)-\nu_{e1}\right),\\
T_2(p_{\sigma L})&=\sum_{\stackrel{e\in{\cal E}}{t(e)=p_{\sigma L}}}
\left(\frac{i}{2}\left((B_e^{RL})^\dagger (B_e^{LR})^\dagger-B_e^{LR} B_e^{RL}\right)-\nu_{e2}\right),\\
T_3(p_{\sigma L})&=\sum_{\stackrel{e\in{\cal E}}{t(e)=p_{\sigma L}}}
\left(\frac{1}{2}\left((B_e^{RL})^\dagger B_e^{RL}-B_e^{LR} (B_e^{LR})^\dagger\right)-\nu_{e3}\right),
\end{align}
and
\begin{align}
T_1(p_{\sigma R})&=\sum_{\stackrel{e\in{\cal E}}{t(e)=p_{\sigma R}}}
\left(\frac{1}{2}\left(B_e^{RL} B_e^{LR}+(B_e^{LR})^\dagger (B_e^{RL})^\dagger\right)-\nu_{e1}\right),\\
T_2(p_{\sigma R})&=\sum_{\stackrel{e\in{\cal E}}{t(e)=p_{\sigma R}}}
\left(\frac{i}{2}\left((B_e^{LR})^\dagger (B_e^{RL})^\dagger-B_e^{RL} B_e^{LR}\right)-\nu_{e2}\right),\\
T_3(p_{\sigma R})&=\sum_{\stackrel{e\in{\cal E}}{t(e)=p_{\sigma R}}}
\left(\frac{1}{2}\left(B_e^{RL} (B_e^{RL})^\dagger-(B_e^{LR})^\dagger B_e^{LR}\right)-\nu_{e3}\right).
\end{align}

\subsubsection{Operator Form}\label{DiracForm}
There is yet another way of presenting the moment map expressions.  Let us introduce the Dirac operator
\begin{align}\label{DiracOperator}
{\cal D}=\left(\begin{array}{c}    
-\frac{d}{ds}+i{\bf T}^*\\
Q^\dagger\\
\left(B^-\right)^\dagger\\
\left(B^+\right)^\dagger
\end{array}\right).
\end{align}
Strictly speaking, this is called the Weyl operator in the physics literature, while the Dirac operator in physics is  
$\left(\begin{array}{cc}
0 & {\cal D}\\
{\cal D}^\dagger & 0
\end{array}\right),$ with ${\cal D}$ given by Eq.~\eqref{DiracOperator}.
From the expression \eqref{DiracOperator} it is clear that 
${\cal D}: \Gamma(S\otimes E)\rightarrow \Gamma(S\otimes E)\oplus W\oplus E_L\oplus E_R.$

Its hermitian conjugate operator is
\begin{align}
{\cal D^\dagger}=\frac{d}{ds}-i{\bf T}+\sum_{\lambda\in\Lambda_0}\delta(s-\lambda)Q_\lambda
+\sum_{e\in{\cal E}}\left(\delta(s-t(e))B_e^-+\delta(s-h(e))B_e^+\right).
\end{align}
Since we can use the hermitian structure to identify the spaces with their dual, we view ${\cal D}^\dagger: \Gamma(S\otimes E)\oplus W\oplus E_L\oplus E_R\rightarrow\Gamma(S\otimes E).$

The moment map then has the form
\begin{equation}
\mu(Q,B,{\bf T})={\rm Im}\, (-i){\cal D}^\dagger{\cal D}.
\end{equation}

\section{The Moduli Space of a Bow Representation and Natural Bundles on It}
\label{Sec:Moduli}
As we discussed in Sec.~\ref{Sec:BowSolution}, the moduli space of a bow representation $\mathfrak R$ is the hyperk\"ahler reduction of ${\rm Dat}\, \mathfrak R$ by the gauge group ${\cal G}$
\begin{equation}
{\cal M}=\mu^{-1}/{\cal G}={\rm Dat}\, {\mathfrak R}/\!\!/\!\!/{\cal G}.
\end{equation}
The resulting space ${\cal M}$ is hyperk\"ahler, thus there is a two-sphere parameterizing the complex structures and we can view it as a complex variety in any one of these complex structures.  For a unit vector $\vec{n}$ the corresponding complex structure is $n_1 e_1+n_2 e_2+n_3 e_3.$ For example for $\vec{n}=(0,0,1)$ the space ${\rm Dat}\, \mathfrak R$ can be viewed as a complex variety parameterized by $(I,J,B^{LR}, B^{RL}, (D, T)),$ where $D=\frac{d}{ds}-i T_0+T_3$ is viewed as a (not necessarily hermitian) connection  on $E\rightarrow{\cal I}.$  
There is a natural action of the complexification of the gauge group ${\cal G}^\C$ on ${\rm Dat}\, \mathfrak R.$  According to the result of \cite{Kronheimer:1988},  the complex symplectic reduction of ${\rm Dat}\, \mathfrak R$ with respect to ${\cal G}^\C$ is isomorphic as a complex manifold to ${\cal M}$ viewed in the corresponding complex structure.  In other words the complex symplectic quotient 
\begin{equation}
{\rm Dat}/\!\!/{\cal G}^\C=(\mu_1+i\mu_2)^{-1}(\nu_1+i\nu_2)/{\cal G}^\C=\mu^{-1}(\nu)/{\cal G}={\cal M}.
\end{equation}
It is an infinite dimensional generalization of a theorem in \cite{Hitchin:1986ea}.  The proof of this statement is essentially that of Donaldson in \cite{Donaldson:1985id}. 

Thus we can view ${\cal M}$ as the space of solutions to the complex moment map condition \eqref{ClxEq} only, modulo the complexified gauge transformations. This is the case for any complex structure one chooses.

\subsection{Natural Bundles on ${\cal M}$}
\label{Sec:NaturalBundles}
The level space $\mu^{-1}(\nu)$ (which is the space of bow solutions satisfying the given moment map conditions) is a subspace of the linear hyperk\"ahler space of all bow data, therefore it has an induced metric.  Moreover, it has a natural isometric action of the gauge group ${\cal G}$.  We defined the moduli space of the bow representation to be 
${\cal M}=\mu^{-1}(\nu)/{\cal G}$ with the quotient metric.  The fact that $\mu^{-1}(\nu)$ carries a ${\cal G}$-invariant metric and that  $\cal M$ is a quotient space implies that there is a natural family of vector bundles with connections on $\cal M.$   In particular, if we choose a point $s'\in{\cal I}$ we can consider the subgroup ${\cal G}_{s'}$ of gauge transformations that act trivially at $s=s',$ i.e. 
\begin{equation}
{\cal G}_{s'}=\left\{g\in{\cal G}\, |\, g(s')=1\right\}.
\end{equation}
Then  the quotient group is the group  $G_{s'}\equiv{\cal G}/{\cal G}_{s'}.$  It can be identified with the group acting on $E_{s'}$ fiber.  Now $R_{s'}=\mu^{-1}(\nu)/{\cal G}_{s'}$ is a finite dimensional space with the quotient metric, moreover, $R_{s'}/G_{s'}={\cal M}.$  If the group action is free we obtain the   principal bundle 
\begin{equation}
\begin{array}{rcl}
G_{s'}&\rightarrow&R_{s'}\\
 &                    &\downarrow\\
 &                    &\cal M
\end{array}
\end{equation}
with a connection determined by the metric on $R_{s'}.$ 

Since we identified the group $G_{s'}$ with the group acting on the fiber $E|_{s'},$ we have an associated hermitian vector bundle  
\begin{equation}
\begin{array}{rcl}
E_{s'}&\rightarrow&R_{s'}\\
 &                    &\downarrow\\
 &                    &\cal M
\end{array}
\end{equation}
Thus we can view ${\cal I}$ as parameterizing natural bundles $R_s\rightarrow{\cal M}$ over the moduli space of the bow representation.  Each $R_s\rightarrow{\cal M}$ carrying a natural connection $d_s.$ The curvature of $d_s$ is self-dual, as in \cite{HitchinDirac}.

\subsection{Holomorphic Description}
In any given complex structure corresponding to some unit vector $\vec{n}$ we view our moduli space as a complex variety ${\cal M}=\left(\mu^{\mathbb{C}}_{\vec{n}}\right)^{-1}(\nu^{\mathbb C}_{\vec{n}})/{\cal G}^{\mathbb C}.$  Here $\mu^{\mathbb{C}}_{\vec{n}}$  and $\nu^{\mathbb C}_{\vec{n}}$ are the complex linear combinations of the components of respectively the moment map and its values determined by the choice of the complex structure $\vec{n}.$  Just as in the real description above $G_{s}^{\mathbb C}={\cal G}^{\mathbb C}/{\cal G}^{\mathbb C}_{s}$ and the variety ${\bf R}_{s}({\vec{n}})\equiv\left(\mu^{\mathbb{C}}_{\vec{n}}\right)^{-1}(\nu^{\mathbb C}_{\vec{n}})/{\cal G}^{\mathbb C}_{s}$ can be viewed as a principal  bundle 
\begin{equation}
\begin{array}{rcl}
G_{s}^{\mathbb C}&\rightarrow&{\bf R}_{s}({\vec{n}})\\
 & &\downarrow\\
 & &\cal M
\end{array}
\end{equation}
with the inherited holomorphic structure.   If we denote by ${\bf E}$ the bundle $E$ without its hermitian structure, then we obtain the associated holomorphic vector bundle
\begin{equation}
\begin{array}{rcl}
{\bf E}_{s}^{\mathbb C}&\rightarrow&{\bf R}_{s}({\vec{n}})\\
 & &\downarrow\\
 & &\cal M.
\end{array}
\end{equation}
Needless to say, these resulting holomorphic bundles do depend on our initial choice of the complex structure determined by $\vec{n}.$

\section{ALF Spaces}
\label{Sec:ALF}
In this section we identify bows and representations that have ALF spaces as their moduli spaces.  These are of either $A$- or $D$-type.  A given ALF space can allow various realizations as a moduli space of different bows.  It will suffice for our purposes to specify for each ALF space some  corresponding bow and its representation that deliver this ALF space as its moduli space.

\subsection{Taub-NUT}
The simplest ALF space of A-type is the Taub-NUT space. It can also be referred to as the $A_0$ ALF space.
\begin{figure}[htbp]
\begin{center}
\includegraphics[width=0.6\textwidth]{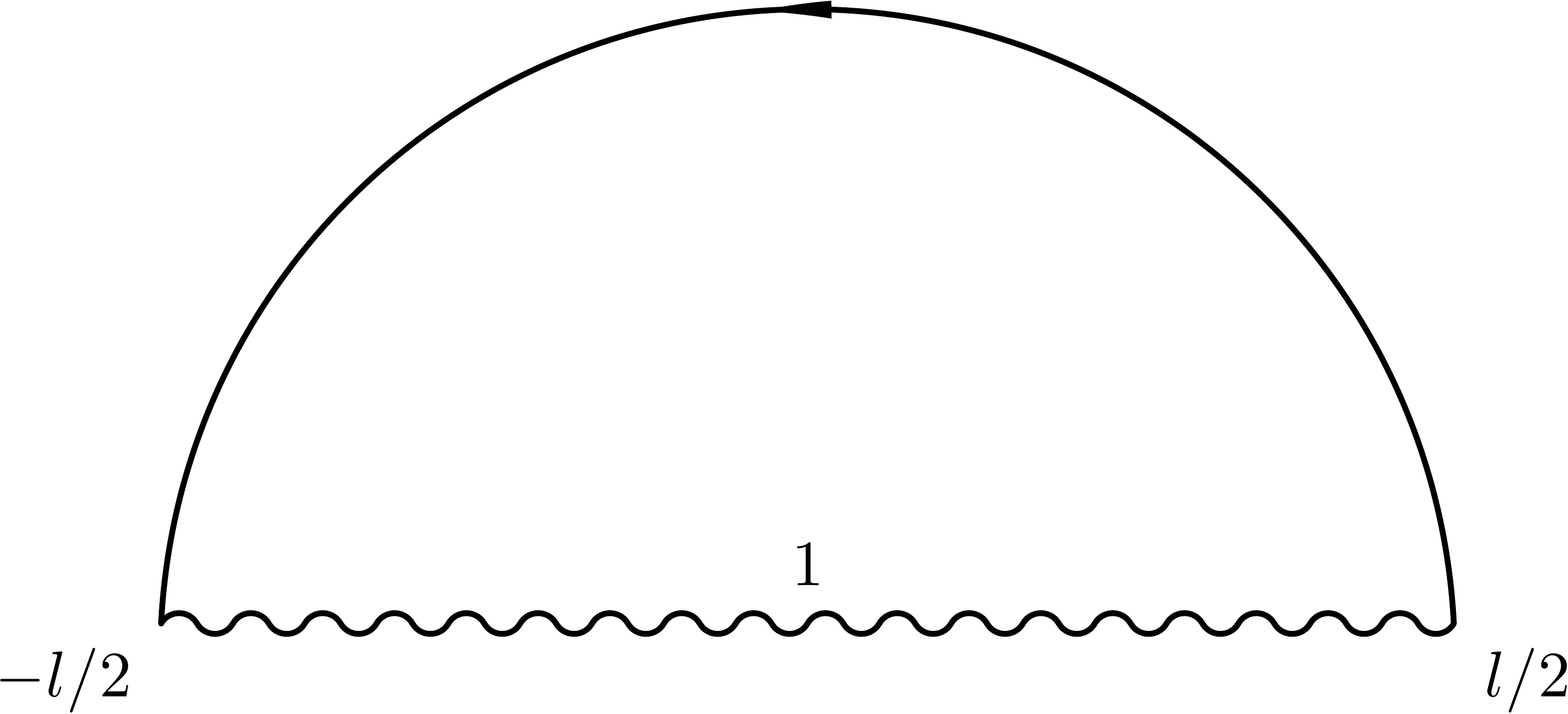}
\caption{A bow consisting of a single interval and a single edge with a representation having a line bundle over the interval.}
\label{TNBow}
\end{center}
\end{figure}
For the rank one Nahm data imposing the moment map conditions in the interior of the interval leads to $\frac{d}{ds}T_j=0,$ thus $T_j$ are constant and a gauge can be chosen so that $T_0$ is also a constant.  Constant $U(1)$ transformation leaves the whole set of bow data inert. Thus the only remaining gauge transformation is the $U(1)$ group acting at $s=l/2.$ Let us denote it by $U_R(1).$  This $U_R(1)$ can be viewed as the factor group of the group of all the gauge transformations modulo the group of gauge transformations that equal to identity at the point $p_R.$  The remaining quotient ${\mathbb R}^3\times S^1\times{\mathbb R}^4/\!\!/\!\!/U_R(1)$ is the Taub-NUT space with the scale parameter given by the length of the interval $l$ and the position of the Taub-NUT center given by negative of $\nu$ if value of the moment map is $\mu=\nu.$

\subsection{Multi-Taub-NUT}
\begin{figure}[htbp]
\begin{center}
\includegraphics[width=0.5\textwidth]{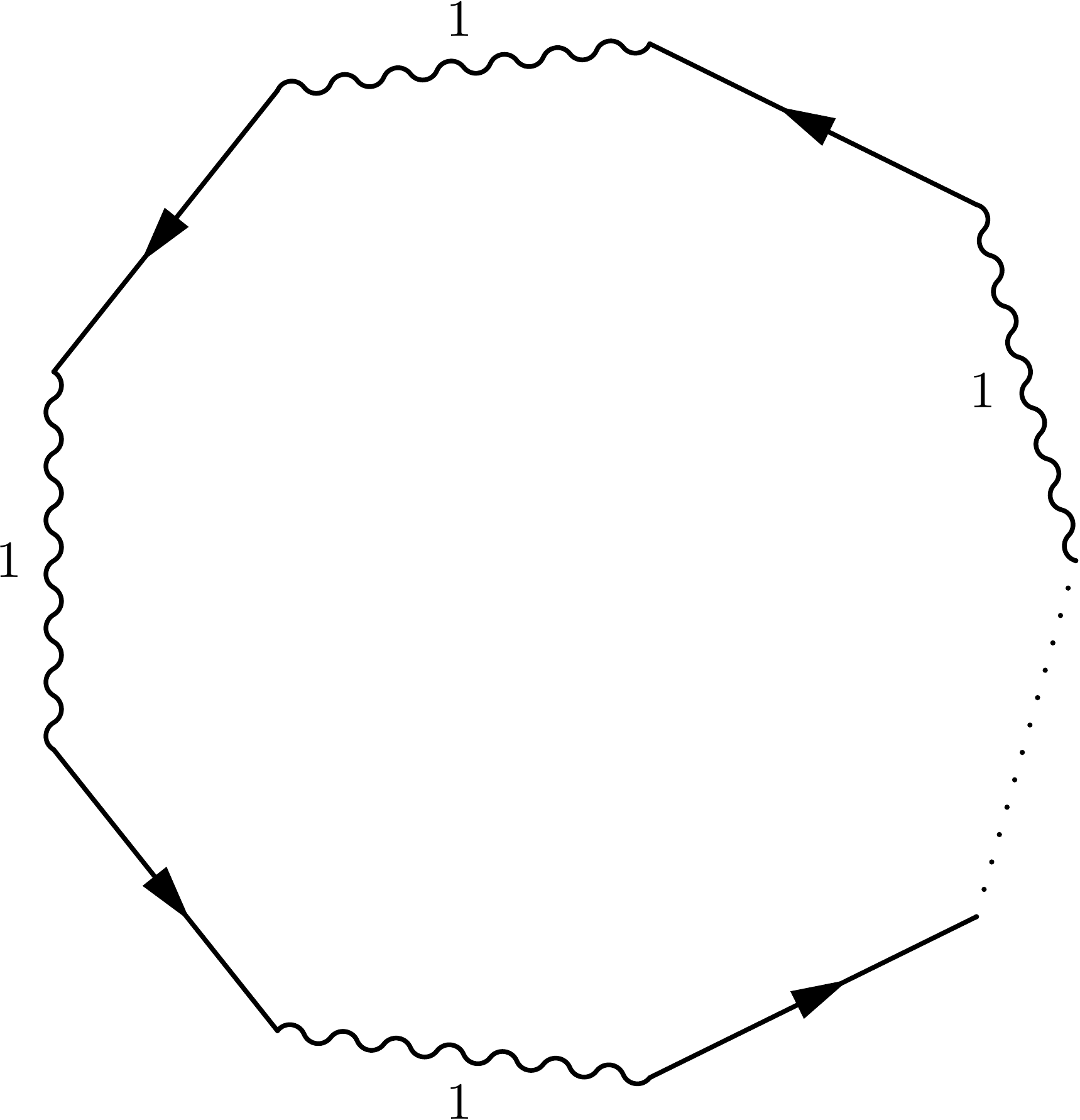}
\caption{For an $A_{k}$ ALF space, also called $(k+1)$-centered Taub-NUT, the bow diagram above contains $k+1$ interval, each carrying a line bundle.}
\label{mTNBow}
\end{center}
\end{figure}

The bow of Figure~\ref {mTNBow} with the representation that consists of line bundles on each interval has the $(k+1)$-centered Taub-NUT space as its moduli space.  Just as in the case of the Taub-NUT above, performing the hyperk\"ahler reduction with respect to the gauge group acting only in the interior of the intervals leads to the space $\big({\mathbb R}^3\times S^1\big)^{k+1}\times\big({\mathbb R}^4\big)^{k+1}$ with the flat metric and sizes of the $S^1$ circles equal to $1/\sqrt{l_\sigma}.$  Performing the quotient with respect to the remaining gauge groups acting at the ends of the intervals is essentially the same calculation as in \cite{Gibbons:1996nt}.  It leads to the multi-Taub-NUT space also called $A_{k}$ ALF space with the scale parameter equal to the total sum of the interval lengths.  The $k+1$ Taub-NUT center positions corresponding to the edges equal to the negative of the value $\nu_e$  that was chosen in the moment map condition $\mu=\sum_{e\in{\cal E}}\big(\delta(s-t(e))-\delta(s-h(e))\big)\nu_e.$

\subsection{$D_k$ ALF Space}\label{Sec:DkALF}
\begin{figure}[htbp]
\begin{center}
\includegraphics[width=0.8\textwidth]{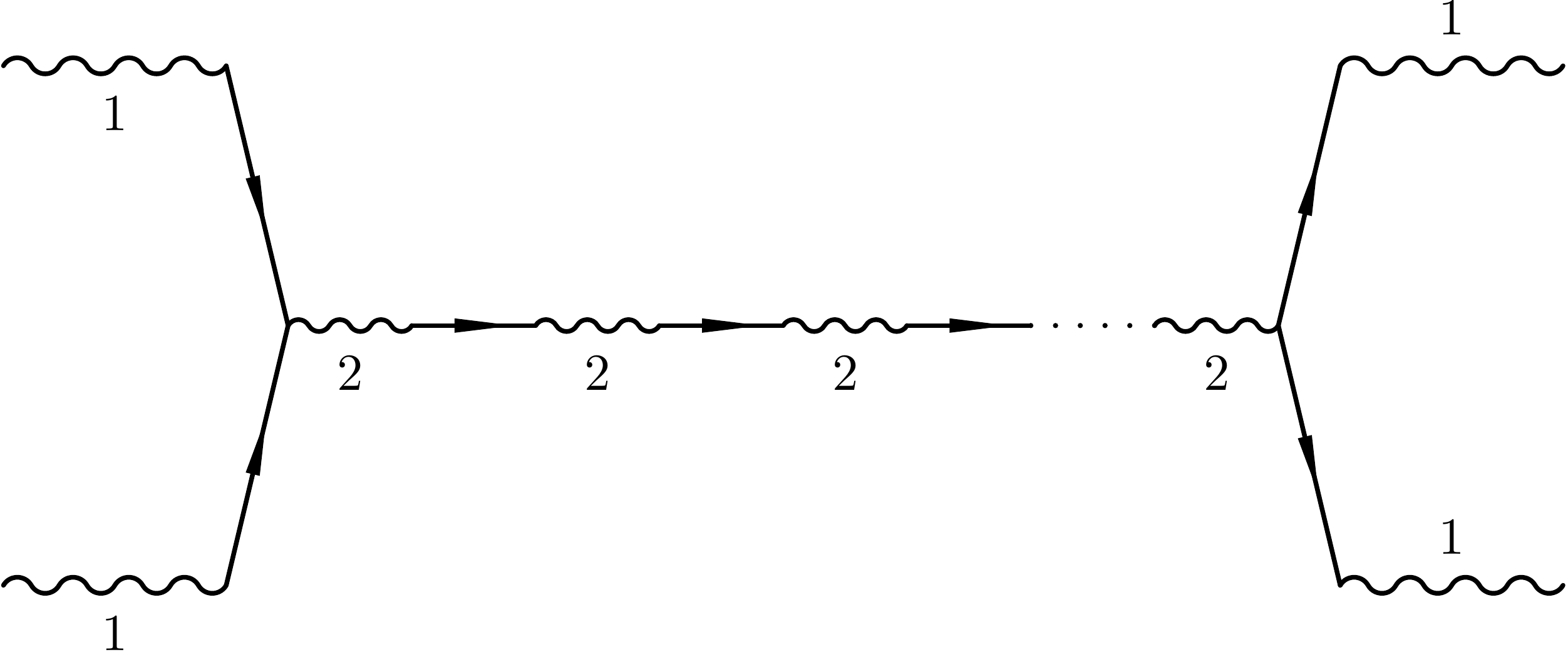}
\caption{We model $D_{k}$ ALF space as the moduli space of this bow representation. This bow has $k-3$ intervals with rank $2$ bundles over them and four intervals with line bundles over them.}
\label{DkALFBow}
\end{center}
\end{figure}

The $D_{k}$ ALE space is the deformation of the quotient of ${\mathbb R}^{4}$ by the action of the dihedral group ${\bf D}_{k-2}$ of order $4k-8.$ According to \cite{Kronheimer:1989zs}, $D_{k}$ ALE space is a moduli space of the affine $D_{k}$ quiver representation determined by the null vector of the corresponding Cartan matrix.  Among the bow deformations of the corresponding quiver, the one with the representation in Fig.~\ref{DkALFBow} has the $D_{k}$ ALF space as its moduli space.

$D_{k}$ ALF space in this form was already considered by Dancer in \cite{Dancer:1992yg}.  Effectively the description of \cite{Dancer:1992yg} is equivalent to a bow which has all but one intervals having zero lengths.

\section{Yang-Mills Instantons on ALF Spaces}\label{Sec:InstBow}
Now that we have modelled each ALF space as a moduli space of a concrete bow representation, say ${\mathfrak s}$ (for `small' representation) we can choose any other representation ${\mathfrak L}$ (for `large' representation') of the same bow.  Any gauge equivalence class of a solution of ${\mathfrak L}$ defines an instanton on the original ALF space up to the action of the gauge group.  Since there are two representations of the same bow involved, let us denote all the components of the solution in the large representation ${\mathfrak L}$ by the capital letters $Q, B,$ and $T$, as in our notation above, and all the components of a solution in the small representation $\mathfrak s$ by the corresponding lower case letters $q$ and $t$.  Note that all the representations we used in the previous chapter to model the ALF spaces were of file zero, i.e. they had no $\lambda$-points and no associated fundamental multiplets.  In choosing the solution one has to be careful to match the levels of the two representations.  If the original ALF space was the moduli space of the small representation ${\mathfrak s}$ at level $\nu,$ i.e. the solutions $(q,t)$ satisfied 
\begin{equation}
\mu_{\mathfrak s}(q,t)=1_e\otimes\nu=1_e\otimes\sum_{e\in{\cal E}}\big(\delta(s-t(e))-\delta(s-h(e))\big) \nu_e,
\end{equation} 
then the solution of the large representation has to be chosen at level $-\nu,$ i.e. it satisfies 
\begin{equation}
\mu_{\mathfrak L}(Q,B,T)=-1_E\otimes\nu=-1_E\otimes\sum_{e\in{\cal E}}\big(\delta(s-t(e))-\delta(s-h(e))\big) \nu_e.
\end{equation} 

The bow representations below give some examples.

\subsection{Examples}
\subsubsection{$U(2)$ Instantons on the Taub-NUT}
\begin{figure}[htbp]
\begin{center}
\includegraphics[width=0.6\textwidth]{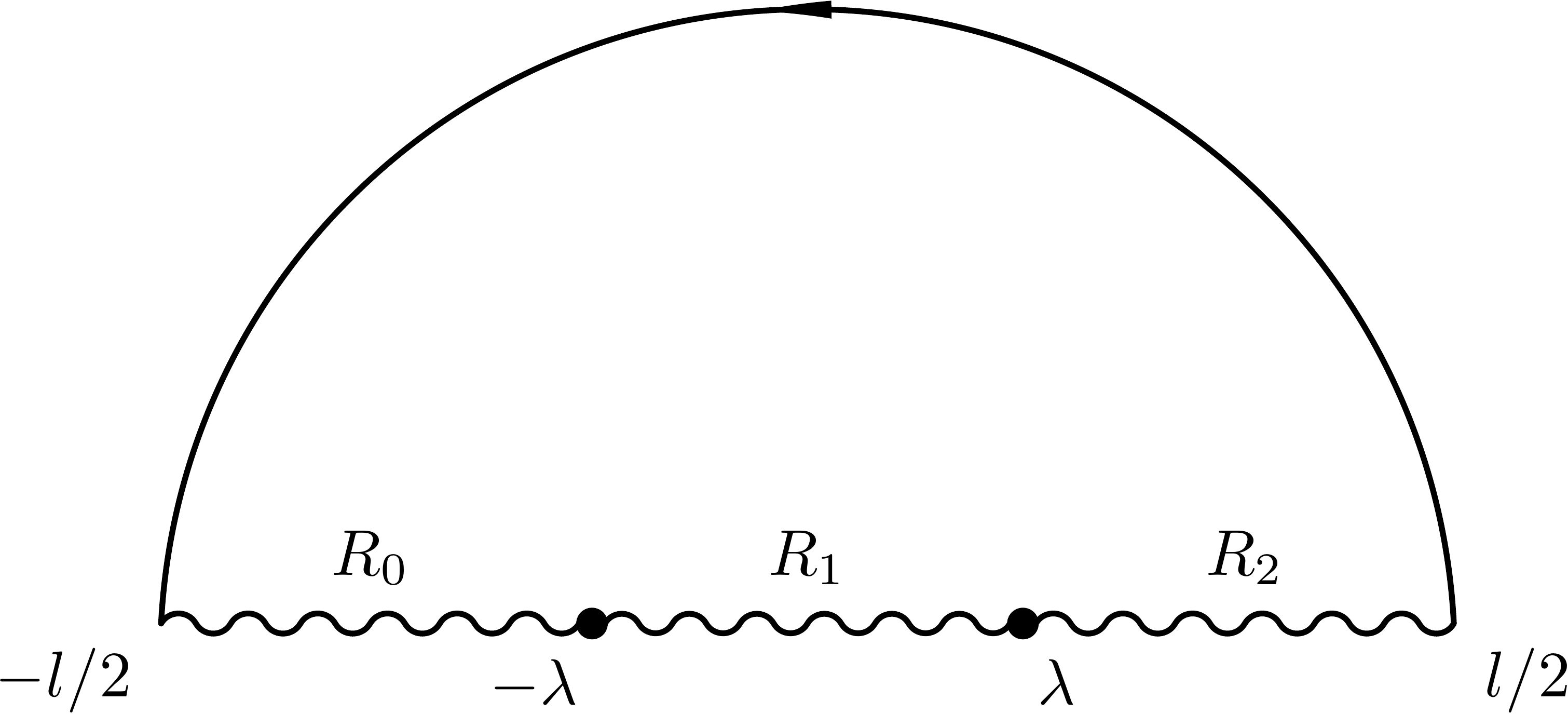}
\caption{A $U(2)$ self-dual connection on Taub-NUT is determined by this bow.  Its instanton number and monopole charges are determined by the ranks $R_{0}, R_{1},$ and $R_{2}$ of the vector bundles.}
\label{SU(2)onTN}
\end{center}
\end{figure}

Figure~\ref{SU(2)onTN} gives a representation of the bow determined by the three nonnegative integer numbers $R_0, R_1,$ and $R_2.$  A solution of this representation determines a $U(2)$ instanton on the Taub-NUT space, with the ranks $R_0, R_1, R_3$ determining its instanton number $m_0$ and its monopole charges. 

\subsubsection{$U(n)$ Instantons on the Taub-NUT}
\begin{figure}[htbp]
\begin{center}
\includegraphics[width=0.7\textwidth]{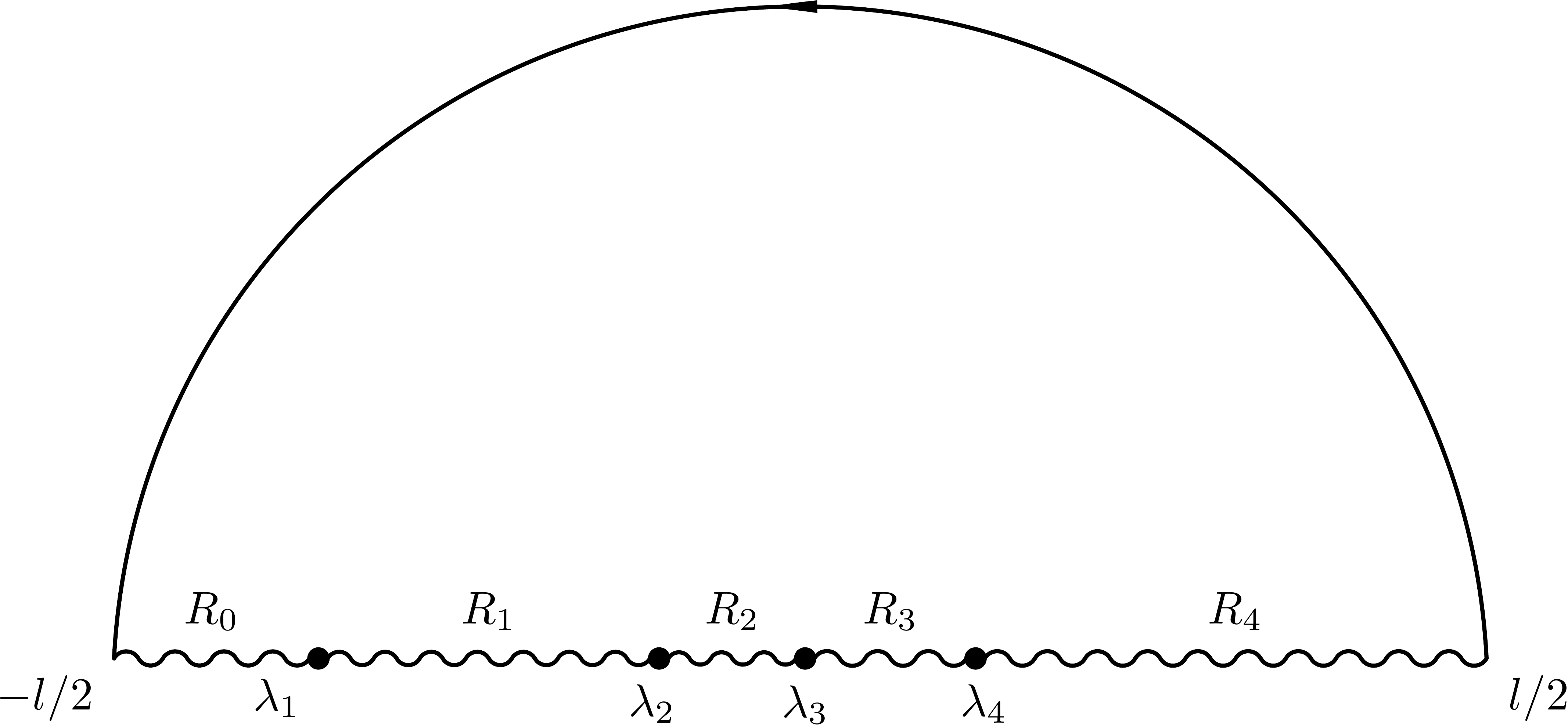}
\caption{A $U(4)$ instantons on a Taub-NUT space are in one-to-one correspondence with the solutions of this bow representation.  $e^{\lambda_{1}}, e^{\lambda_{2}}, e^{\lambda_{3}},e^{\lambda_{4}}$ are the values of the monodromy at infinity, while the vector bundle ranks $R_0, R_{1}, R_{2},R_{3}, R_{4}$ determine the instanton number and its monopole charges.}
\label{U(n)onTN}
\end{center}
\end{figure}

A generic $U(n)$ instanton on the Taub-NUT space is given by a general regular representation of the Taub-NUT bow of Figure~\ref{TNBow} of file $n$, such as the one in Figure~\ref{U(n)onTN}.   The positions of the $\lambda$-points are given by the values of the logarithm of the eigenvalues of the holonomy of the instanton connection around the compact direction of the Taub-NUT triholomorphic isometry at infinity.  While the ranks $R_0^j=R_j$ with $j=0,1,\ldots, n$ can be any nonnegative integers.  These ranks determine the monopole charges and the monopole number.

\subsubsection{$U(n)$ Instantons on the $(k+1)$-centered Taub-NUT}
\begin{figure}[htbp]
\begin{center}
\includegraphics[width=0.5\textwidth]{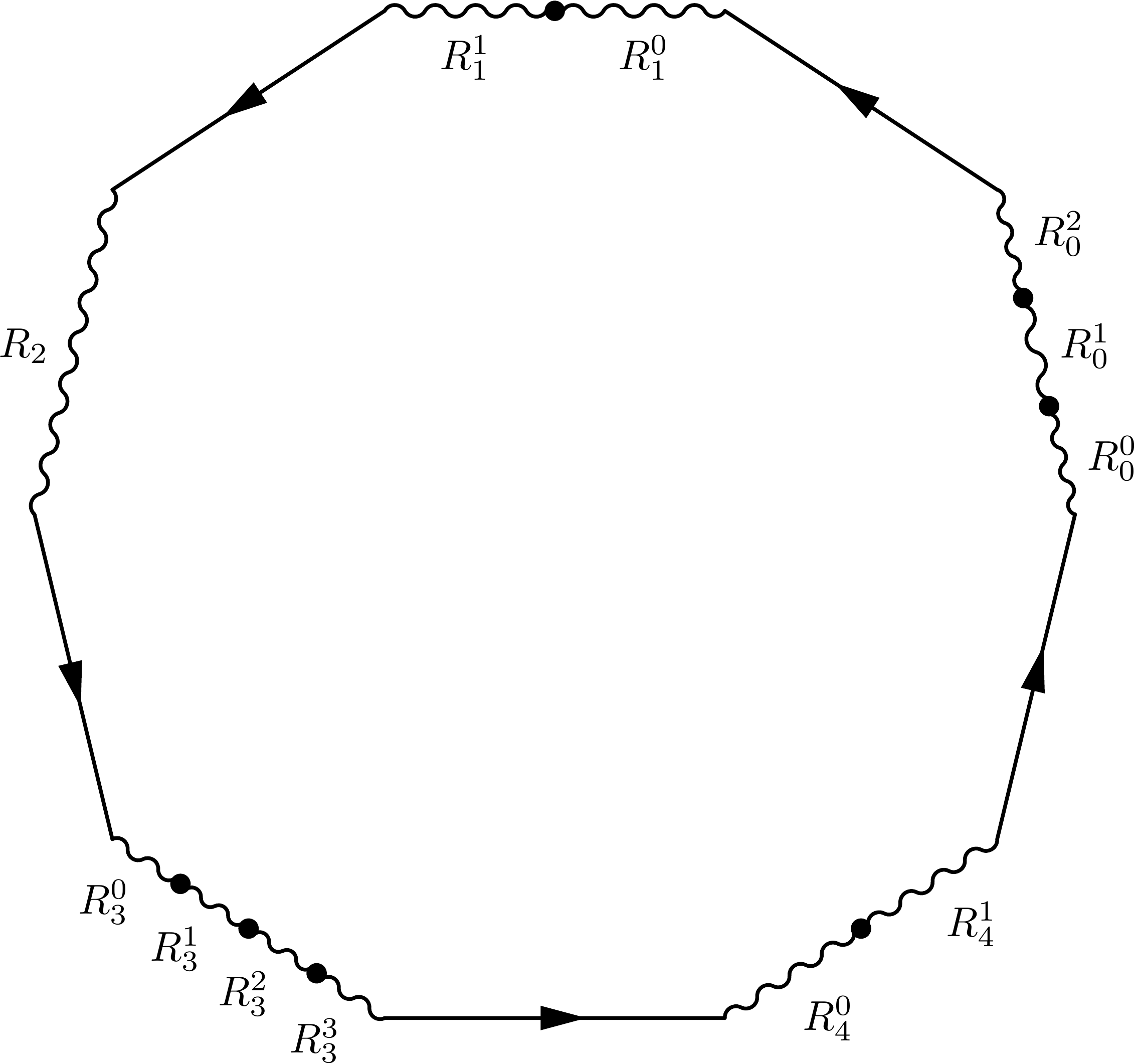}
\caption{A bow representation determining a $U(7)$ instanton on the $5$-centered Taub-NUT.}
\label{U(n)onmTN}
\end{center}
\end{figure}

A generic $U(n)$ instanton on a $(k+1)$-centered Taub-NUT space is given by a regular file $n$ representation of the $A_{k}$ bow of Figure~\ref{mTNBow}, such as the file seven representation given in Figure~\ref{U(n)onmTN} giving a $U(7)$ instanton on $A_4$ ALF space.  

\subsubsection{$U(n)$ Instantons on the $D_{k}$ ALF Space}\label{Sec:DkALFInst}
\begin{figure}[htbp]
\begin{center}
\includegraphics[width=0.8\textwidth]{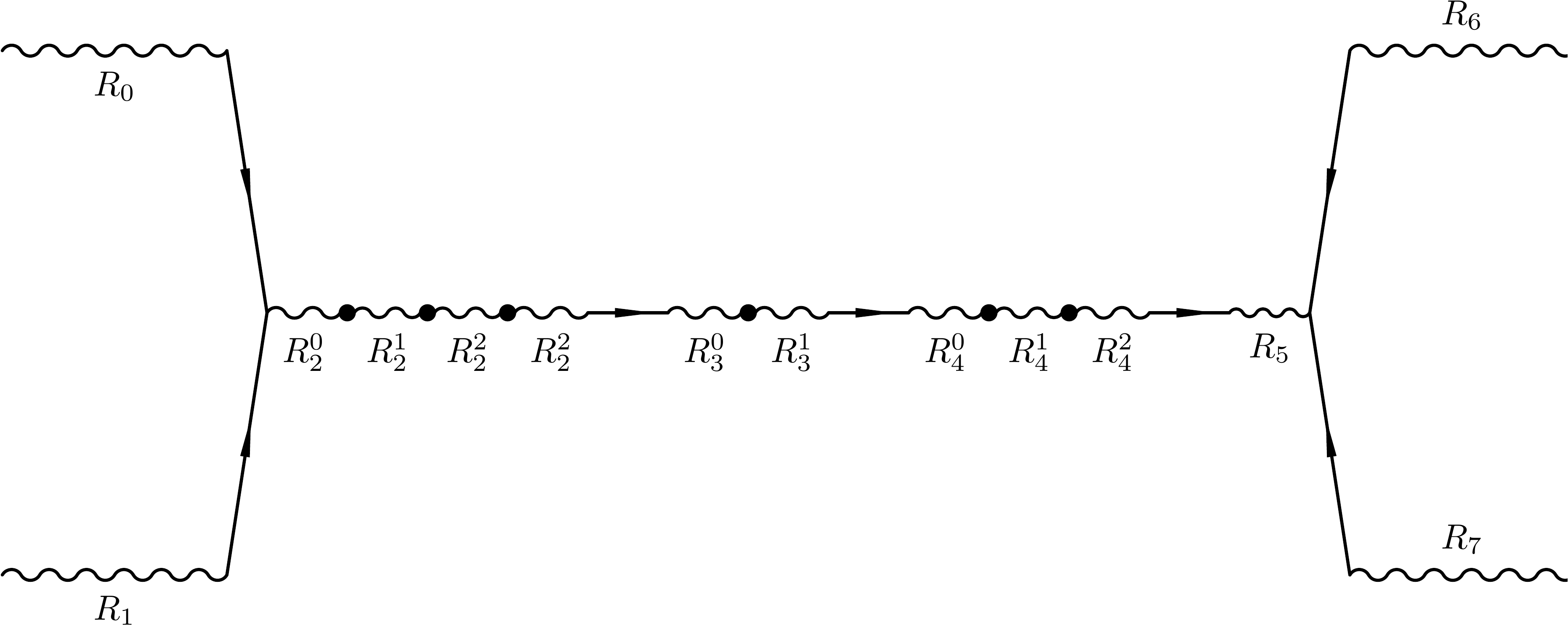}
\caption{An example of a $U(6)$ instanton on the $D_{7}$ ALF space.}
\label{U(n)onDkALF}
\end{center}
\end{figure}

A generic $U(n)$ instanton on the $D_k$ ALF space is given by a regular file $n$ representation of the bow in Figure~\ref{DkALFBow}, such as the one in Figure~\ref{U(n)onDkALF}.

\subsection{Instanton Charges}\label{Sec:Charges}
As we demonstrate in Sections \ref{Sec:NahmTransform} and \ref{Sec:Cohomology}, if a bow has a representation $\mathfrak s$ with a four-dimensional moduli space ${\cal M}_{\mathfrak s}(\nu)$ for some moment map value $\nu$, then 
for any other bow representation $\mathfrak L$ of file $n$ its solution with the moment map value $-\nu$ produces an $U(n)$ instanton on ${\cal M}_{\mathfrak s}(\nu).$  We would like to understand the relation between the representation and the topological charges of the resulting instantons.  Let us focus from now on on the case of instantons on a multi-Taub-NUT space.  A Taub-NUT with $k+1$-centers is a moduli space of the $A_{k}-$bow representation of Figure~\ref{mTNBow}.  This $A_{k}$ ALF space is often denoted by $TN_{k+1}.$

There are three kinds of topological charges one can associate to a rank $n$ finite action anti-self-dual connection on an $A_{k}$ ALF space:
\begin{itemize}
\item $n$ nonnegative integer monopole charges $(j^1, j^2,\ldots, j^n)$,
\item $k+1$ first Chern class values $c_1$,
\item the value of the second Chern class $c_2$ we denote by $m^0.$
\end{itemize}
The Chern classes were computed by Witten in \cite{Witten:2009xu}.
In order to discuss these charges we have to recall a few facts about the $(k+1)-$centered Taub-NUT space.  

It is convenient to associate to an $A_{k}$ bow representation a circle of length $l=l_0+l_1+\ldots+l_k$ parameterized by a coordinate $s$ with $k+1$ points $0\leq p_\sigma<l,\ \sigma=0,\ldots,k$ with $p_\sigma$ positioned at $s=\sum_{\varsigma=0}^\sigma l_{\varsigma}.$  We understand the interval $I_\sigma$ of the bow to be associated to the arc $[p_\sigma, p_{\sigma+1}].$   For a representation of file $n$ this assignment gives a natural position for the $n$ points $\lambda_\sigma^\alpha$ within these arcs on the circle.  Let the coordinates of these points in the same order as they appear along the circle be $\lambda_\tau,\ \tau=1,\ldots,n.$  This way the $\lambda$-points are labelled in two ways: as $\lambda_\sigma^\alpha$ with $\sigma=0,\ldots, k$ and $\alpha=1, \ldots, r_\sigma$ and as $\lambda_\tau$ with $\tau=1,\ldots, n.$  This gives maps $\sigma(\tau)$ and $\alpha(\tau)$ such that $\lambda_\tau=\lambda_{\sigma(\tau)}^{\alpha(\tau)}.$  Now the circle is divided into subarcs by the points $p_\sigma$ with $\sigma=0\ldots,k$ and $\lambda_\tau$ with $\tau=1,\ldots,n.$  Each of these subarcs is associated to a subinterval of the bow representation which carries a Hermitian bundle over it.  We assign the rank of this bundle $R_\sigma^\alpha$ to each subarc. 

We model the $TN_k$ as a moduli space of the representation of Figure \ref{mTNBow} with the moment map $\mu=\sum_e (\delta(s-t(e))-\delta(s-h(e))\nu_e$ for some given set of distinct $k$ three vectors $\vec{\nu}_0,\ldots, \vec{\nu}_k$ with $\nu_e=e_1\otimes\nu_e^1+e_2\otimes\nu_e^2  +e_3\otimes\nu_e^3.$  The hyperk\"ahler quotient in this setup was considered in \cite{Gibbons:1996nt} and leads to the following metric
\begin{equation}
ds^2=\frac{1}{4}\left(Vd\vec{r}^{\, 2}+\frac{1}{V}(d\theta+\omega)^2  \right),
\end{equation}
with $\theta\sim \theta+4\pi, V=l+\sum_{\sigma=0}^k\frac{1}{r_\sigma},$ where $r_\sigma=|\vec{r}-\vec{\nu}_\sigma|$ and the one-form $\omega$ satisfies $d\omega=*_3 d V.$ Here $*_3$ signifies the Hodge star operation acting on differential forms on ${\mathbb R}^3.$

The hyperk\"ahler reduction produces the following one-forms on this space
\begin{align}
\tilde{a}&=\frac{1}{2}\frac{d\theta+\omega}{V},& 
a^{(\sigma)}=\frac{1}{2}\left(\frac{1}{r_\sigma}\frac{d\theta+\omega}{V}-\eta_\sigma\right),
\end{align}
where $\eta_\sigma$ is a one-form on ${\mathbb R}^3$ satisfying $d\eta_\sigma=*_3 d\frac{1}{r_\sigma}.$ 
The self-dual connection on the natural bundles $R_s$ defined in Section~\ref{Sec:NaturalBundles} can be written in terms of these forms as
\begin{equation}
a_s=s\, \tilde{a}+\sum_{\substack{\sigma\\ l_1+l_2+\ldots+l_\sigma<s}} a^{(\sigma)}.
\end{equation}
This has the form $a=\frac{H}{V}(d\theta+\omega)-\eta$ with $d\eta=*_3 dH.$  For any such form  
\begin{equation}
f=da=-\left((d\tau+\omega)\wedge d\frac{H}{V}+V*_3d\frac{H}{V}\right),
\end{equation}
and 
\begin{equation}
f\wedge f=(d\tau+\omega)\wedge d\frac{H}{V}\wedge\left(d\eta-\frac{H}{V}d\omega\right),
\end{equation}
and $\int_M f\wedge f=\int_{\partial M}\frac{H}{V}\left(d\eta-\frac{H}{V}d\omega\right).$ 
For $H=\frac{1}{2}\left(s+\sum\frac{n_\sigma}{r_\sigma}\right)$ it is clear that the only contribution arises from the integral over the sphere at infinity giving
\begin{equation}
\frac{1}{8\pi^2}\int_{TN_{k+1}}\displaylimits f\wedge f=\frac{1}{2}\left(\frac{s}{l}\right)^2 k-\frac{s}{l}\sum_e n_e.
\end{equation}
This is so since $\lim_{\vec{r}\rightarrow\vec{\nu}_e}\frac{H}{V}=n_e$ and the contribution of $\eta$ to the integral over a small sphere surrounding $\vec{\nu}_e$ is cancelled by that of $\frac{H}{V}\omega.$

The simplest representation of a bow is {\em a skyscraper} or zero-rank representation.  If it is of file $n$ it consists of $n$ $\lambda$-points and has all vector bundles of rank zero.  For such a representation the connection is
$A={\rm diag} (a_{\lambda_1}, a_{\lambda_2},\ldots, a_{\lambda_n})$
and 
\begin{equation}
\frac{1}{8\pi^2}\int_{TN_{k+1}}\displaylimits{\rm tr}\, F\wedge F=\frac{k+1}{2}\sum_{\tau=1}^n \left(\frac{\lambda^\tau}{l}\right)^2-\sum_{\tau=1}^n\sum_{\substack{\sigma=0 \\ p_\sigma<\lambda_\tau}}^k \frac{\lambda^\tau}{l}.
\end{equation}

Let us specify the monopole charges.  Consider a holonomy along the isometric direction at infinity parameterized by $\theta.$  As this direction is finite for all ALF spaces (see \cite{Minerbe}) the isometry has eigenvalues that are unrestricted. We presume the asymptotic monodromy to be generic and thus all of these eigenvalues to be distinct and numbered respecting their cyclic order.  Asymptotically the logarithm of these eigenvalues does not depend on the direction in which one moves towards infinity and behaves as 
\begin{equation}
\tau^{\rm{}th} {\rm eigenvalue}= \frac{2\pi i}{V}\left(\lambda^\tau+j^\tau/r+O(1/r^2)\right),
\end{equation}
where all the integers $j^\tau$ can be chosen to satisfy $0\leq j^\tau<k.$ 
A geometric description of the integers $j^\tau,  \tau=1,2,\ldots, n$ is that these are the Chern numbers of the holonomy eigen-bundles over the sphere of directions in the base space.  These are defined modulo $k.$
The reason these are associated with monopole charges is that any finite action connection $A$ asymptotically approaches the connection 
\begin{equation}\label{Eq:app}
A_{\rm app}=g^{-1}\left(A_3+\Phi\frac{d\theta+\omega}{V}\right)g+i g^{-1}dg,
\end{equation}
with $A_3$ some connection over a bundles over ${\mathbb R}^3$ and $\Phi$ an endomorphism of that bundle.  The self-duality condition on $A$ implies \cite{KronheimerMS} that the asymptotic fields $A_3$ and $\Phi$ satisfy the monopole equation of Bogomolny
\begin{equation}
F_3=*_3 D_{A_3}\Phi.
\end{equation}
Here $F_3$ is the curvature of the connection $A_3,$ $D_{A_3}$ is the covariant differential with the $A_3$ connection, and $*_3$ is the Hodge star operation acting on $T^*{\mathbb R}^3.$  The conventional monopole charges are exactly the quantities $j^\tau$ that we defined above.

Viewing the $TN_{k+1}$ as a circle fibration over ${\mathbb R}^3\setminus\{\nu_\sigma\},$ let us pick a set of $k+1$ semi-infinite nonintersecting lines, such that the $\sigma^{\rm th}$ line originates at $\nu_\sigma.$  The preimage in the total $TN_{k+1}$ of this line is an infinite cycle which we denote as $C_\sigma.$  The $k+1$ values of the first Chern class are given by 
\begin{equation}
c_1^\sigma=\frac{1}{2\pi}\int_{C_\sigma} {\rm tr}\, F.
\end{equation}
These are not generally integers, however, a difference of any two of these is an integer.

From the argument in \cite{Witten:2009xu} it follows that the monopole charges and the first Chern class values are given in terms of the bow ranks by
\begin{align}
j^\tau&=R_{\sigma(\tau)}^{\alpha(\tau)+1}-R_{\sigma(\tau)}^{\alpha(\tau)}+\sum_{\substack{\sigma=0 \\ p_\sigma<\lambda_\tau} }^k1,\\
c_1^\sigma&=\sum_\tau \lambda_\tau+R_{\sigma(\tau)}^0-R_{\sigma(\tau)}^{r_\sigma(\tau)}-\sum_{\substack{\tau=1 \\ p_\sigma<\lambda_\tau}}^n 1.
\end{align}

The instanton number is typically defined by the value of the second Chern class $\frac{1}{4\pi^2}\int {\rm tr}\, F\wedge F.$ For compact manifolds this is an integer equal to the instanton number.  In the case of the noncompact base there is no reason this is an integer.  This number receives contributions from the instantons inside, which is integer for a smooth base space, and the monopole contributions.  To single out the instanton contribution consider a large ball $B_R$ of radius $R.$  Inside the ball we have the smooth connection $A_{\rm in}$ and outside the ball we have the connection $A_{\rm out}$ approaching $A_3+\Phi\frac{d\theta+\omega}{V}$ and the two are related by a gauge transformation $g$ as Eq.~\eqref{Eq:app}: $A_{\rm in}=g^{-1}A_{\rm out}g+ig^{-1}dg.$  The second Chern class is a differential of the Chern-Simons form, ${\rm tr}\, F\wedge F=d CS,$ with $CS={\rm tr}\,  A\wedge dA+\frac{2}{3} A\wedge A\wedge A$ and the difference $CS_{\rm in}-CS_{\rm out}=\frac{1}{3} {\rm tr}\,  g^{-1}dg\wedge g^{-1}dg\wedge g^{-1}dg.$  Combining these observation we have 
\begin{align}
\int_{TN_k} {\rm tr}\,  F\wedge F&=\int_{B_R}{\rm tr}\,  F\wedge F+\int_{TN_k\setminus B_R}{\rm tr}\,  F\wedge F\\
 &=\int_{\partial B_R} \frac{1}{3} {\rm tr}\,  g^{-1}dg\wedge g^{-1}dg\wedge g^{-1}dg+\int_{S^3_\infty/{\mathbb Z}_k} CS_{\rm out}.
\end{align}
The last term can be reexpressed in terms of the monopole charges.  This suggests to define the instanton number to be 
\begin{equation}
m_0=\frac{1}{12\pi^2} {\rm tr}\,  g^{-1}dg\wedge g^{-1}dg\wedge g^{-1}dg.
\end{equation}
We expect it to be given in terms of the minimal rank ${\rm min} \{ R_\sigma^\alpha \}$ for the balanced representations we define in Section~\ref{Sec:Balanced}, while for the general representation we expect if to be equal to the minimal integer in the set $\{M_\tau\}$ defined in Section~\ref{Sec:GenRep}.

\section{Nahm Transform}\label{Sec:NahmTransform}

Consider a pair $\mathfrak L$ and $\mathfrak s$ of representations of the same bow.  As in Section~\ref{Sec:InstBow}, we shall refer to the corresponding representations, Nahm data, and solutions as large and small respectively.  To distinguish the ingredients of these two representations we use $E$ and  $W$ to denote the collection of hermitian vector bundles and auxiliary spaces defining the large representation ${\mathfrak L},$ and $e$ and $w$ to denote the collection of  bundles and auxiliary spaces (if any) defining the small representation ${\mathfrak s}.$  We use $E_{L}, E_{R}, e_{L},$ and $e_{R}$ to denote the collections of fibers of the respective bundles at the left and right ends of the intervals of $\cal I.$  For any maps, such as $T_{j}, B_{\pm}$ or $Q$ acting on $E$ or $W$ we use capital letters.  Analogous maps, such as respectively $t_{j}$ or $b_{\pm}$ acting on $e$ or $w,$ will be denoted by lower case letters.   If $\mu_{\mathfrak L}$ is the moment map for the data in the $\mathfrak L$ representation and $\mu_{\mathfrak s}$ for those in the $\mathfrak s$ representation, then we chose our moment map conditions to be
\begin{align}\label{Matching}
\mu_{\mathfrak L}&=-\nu &&{\rm and}&\mu_{\mathfrak s}=+\nu.
\end{align}
 Relying on the form of the moment map of Section~\ref{DiracForm}, one can introduce the Dirac operator ${\cal D}_{\mathfrak L}$ using a bow solution in  ${\mathfrak L}$ as in Eq.~\eqref{DiracOperator}.  In terms of this operator  $\mu_{\mathfrak L}={\rm Im}\, i {\cal D}_{\mathfrak L}^\dagger {\cal D}_{\mathfrak L}.$  Similarly, if ${\cal D}_{\mathfrak s}$ is the Dirac operator for the data in ${\mathfrak s},$ the moment map $\mu_{\mathfrak s}={\rm Im}\, i {\cal D}_{\mathfrak s}^\dagger {\cal D}_{\mathfrak s}.$
 
Now let us consider a `twisted' Dirac operator
 \begin{equation}
{\cal D}_t={\cal D}_{\mathfrak L}\otimes 1_e+1_E\otimes{\cal D}_{\mathfrak s}.
\end{equation}
It acts on $\Gamma(S\otimes E\otimes e)\oplus W\oplus w\oplus 
\left(E_{L}\otimes_{{\cal E}}e_{R}\right)\oplus \left(E_{R}\otimes_{{\cal E}} e_{L}\right).$

A crucial observation for what follows is that, due to our choice of the moment map conditions in Eq.~\eqref{Matching}, the twisted Dirac operator in purely real:
\begin{equation}
{\rm Im}\, {\cal D}_t^\dagger {\cal D}_t=0.
\end{equation}
Moreover, it is strictly positive away from the degenerate locus $D$ in the direct product $\left(\mu_{\mathfrak L}\right)^{-1}(-\nu)\times\left(\mu_{\mathfrak s}\right)^{-1}(\nu).$
In fact, in all of the examples we consider for a generic point ${\rm pt}\in\left(\mu_{\mathfrak L}\right)^{-1}(-\nu)$ the degenerate locus $D$ does not intersect ${\rm pt}\times\left(\mu_{\mathfrak s}\right)^{-1}(\nu).$ This is so since the operator ${\cal D}_t^\dagger {\cal D}_t$ is a sum of a number of nonnegative parts of the form $(T_j\otimes 1_e+1_E\otimes t_j)^2.$  If all of these parts have a zero eigenvalue, then each $T_j$ has this eigenvalue within each interval $I_\sigma,$ which is not true for a generic solution.  This degeneration corresponds to the zero-size limit of an instanton. 
In particular, the positivity implies that ${\rm Ker}\, {\cal D}$ is empty.  Presuming ${\cal D}$ is Fredholm, which is also generically the case,
the vector space ${\rm Ker}\, {\cal D}^\dagger$ is finite dimensional.  If we fix some bow solution of $\mu_{\mathfrak L}=-\nu$ we obtain a vector bundle 
\begin{equation}
{\rm Ker}\, {\cal D}_t^\dagger\rightarrow \mu_{\mathfrak s}^{-1}(\nu).
\end{equation}
In order to understand it better, let us first consider the space of sections of $e\rightarrow {\cal I}.$ We can view it as a fiber of a trivial infinite-dimensional hermitian vector bundle over $\left(\mu_{\mathfrak s}\right)^{-1}(\nu):$
\begin{equation}
\begin{array}{ccc}
\Gamma(e\rightarrow {\cal I})&\rightarrow&{\cal B}\\
 & &\downarrow\\
 & &\mu_{\mathfrak s}^{-1}(\nu)
\end{array}
\end{equation}
The gauge group action is such that the gauge group ${\cal G}$ acts simultaneously on the base $\left(\mu_{\mathfrak s}\right)^{-1}(\nu)$ and maps the corresponding fibers of ${\cal B}$ into each other.  
\begin{equation}
{\cal G}\lefttorightarrow
\begin{array}{c}
{\cal B}\\
\downarrow\\
\mu^{-1}(\nu)
\end{array}
\end{equation}
Thus using this action to identify the fibers locally along the gauge group orbits in the base, we obtain the pushdown bundle $\tilde{{\cal B}}\rightarrow {\cal M}(\nu)=\mu_{\mathfrak s}(\mu).$
This bundle has a nontrivial connection $\nabla_e.$

Turning to the bundle ${\rm Ker}\, {\cal D}_t$ we can view it as a subbundle of the trivial bundle with the fiber given by the space of sections of $S\otimes E\otimes e\rightarrow \mu^{-1}(\nu).$   The trivial connection  induces a connection on the subbundle ${\rm Ker}\, {\cal D}_t\rightarrow {\cal M}(\nu)$. Trivializing ${\rm Ker}\, {\cal D}_t$ along the orbits of the gauge group ${\cal G}$ acting on $e$ this connection descends to a connection on ${\cal M}=\left(\mu_{\mathfrak s}\right)^{-1}(\nu)/{\cal G}.$  In the next section we will prove that this induced connection is self-dual.  To demonstrate this we show that in any given complex structure on the hyperk\"ahler base ${\cal M}(\nu)$ the curvature of this connection is of type $(1,1).$  This is equivalent to the corresponding holomorphic bundle being flat in every complex structure.

\section{Cohomological Interpretation of the Nahm Transform}\label{Sec:Cohomology}
The space of all complex structures forms a sphere.  If we parameterize this sphere by a unit tree-vector $\vec{n},$ in the representation \eqref{Eq:Quaternions} of Section \ref{Sec:HKStructure}, the action of the corresponding complex structure on $S$ is given by a unit quaternion
\begin{equation}
n\equiv n_{1}e_{1}+n_{2}e_{2}+n_{3}e_{3}=\frac{i}{\sqrt{1+\zeta\bar{\zeta}}}
\left(\begin{array}{cc}
-1 & -\bar{\zeta}\\
-\zeta & 1
\end{array}\right).
\end{equation}
The complex coordinate $\zeta=(n_{1}+i n_{2})/n_{3}$ parameterizes the Riemann sphere.

Adopting the standard Hodge theory one can identify the kernel of a Dirac operator with a middle cohomology of a complex.  

To simplify our notation in dealing with the twisted Dirac operator, let us introduce 
\begin{align}Z&\equiv T\otimes 1_{e}+1_{E}\otimes t=(T_{1}+i T_{2})\otimes 1_{e}+1_{E}\otimes(t_{1}+i t_{2}),\\
\text{and}&\\
D_{t}&\equiv D\otimes 1_{e}+1_{E}\otimes d\equiv\frac{d}{ds}-i(T_{0}\otimes 1_{e}+1_{E}\otimes t_{0})+(T_{3}\otimes 1_{e}+1_{E}\otimes t_{3}).
\end{align}
Up to an unimportant scalar factor $n{\cal D}_{t}^{\dagger}$ equals to
\begin{multline}\label{nDd}
\left(\begin{array}{cc}
-1 & -\bar{\zeta}\\
-\zeta & 1
\end{array}\right){\cal D}_{t}^\dagger=
\left(\begin{array}{cc}
(D_{t}+\zeta Z^\dagger)^\dagger & (Z-\zeta D_{t}^\dagger)^\dagger\\
-Z+\zeta D_{t}^\dagger & D_{t}+\zeta Z^\dagger
\end{array}\right)
+\sum_{\lambda\in\Lambda_0}\left(\begin{array}{c}  
-(J+\zeta I^\dagger)^\dagger\\
I-\zeta J^\dagger
 \end{array}\right)\\
+\sum_{e\in{\cal E}}\left\{
\left(\delta(s-t(e))
\left(\begin{array}{c}  
(-B^{LR}+\zeta (B^{RL})^\dagger)^\dagger\\
-B^{RL}-\zeta(B^{LR})^\dagger
  \end{array}\right)
+\delta(s-h(e))
  \left(\begin{array}{c}  
 -(b^{RL}+\zeta (b^{LR})^\dagger)^\dagger\\
 b^{LR}-\zeta (b^{RL})^\dagger
    \end{array}\right)\right)\right.\\
\left.
+\left(\delta(s-t(e))
\left(\begin{array}{c}  
(-b^{LR}+\zeta (b^{RL})^\dagger)^\dagger\\
-b^{RL}-\zeta(b^{LR})^\dagger
  \end{array}\right)
  +\delta(s-h(e))
  \left(\begin{array}{c}  
 -(B^{RL}+\zeta (B^{LR})^\dagger)^\dagger\\
 B^{LR}-\zeta (B^{RL})^\dagger
    \end{array}\right)\right)
    \right\}.
\end{multline}
Here the operators on the right-hand side in the first line act on $\Gamma(S\otimes E\otimes e)\oplus W,$
the operators in the second line act on 
 \begin{equation}
E_L\otimes_{\cal E} e_R\equiv\{E_{h(e)}\otimes e_{t{e}} | e\in{\cal E}\},
\end{equation}
and the operators in the third line act on 
 \begin{equation}
E_R\otimes_{\cal E} e_L\equiv\{E_{t(e)}\otimes e_{h{e}} | e\in{\cal E}\},
\end{equation}
so that $n{\cal D}_t^\dagger$ acts on the direct sum of these spaces $\Gamma(S\otimes E\otimes e)\oplus W\oplus \left(E_L\otimes_{\cal E} e_R\right)\oplus(E_R\otimes_{\cal E} e_L).$

Let us consider spaces ${\cal A}^0=\Gamma(S\otimes E\otimes e), {\cal A}^1=\Gamma(S\otimes E\otimes e)\oplus W\oplus (E_L\otimes e_R)\oplus (E_R\otimes e_L),$ and ${\cal A}^2=\Gamma'(S\otimes E\otimes e).$ The latter space consists of distributions and $\Gamma'$ is denoting the fact that the sections we consider have the form $f(s)+\sum_{\lambda\in\Lambda_0}\delta(s-\lambda) a_\lambda.$
Now we can consider a  Dolbeault-type family of complexes which depend on  $\zeta$ holomorphically 
\begin{equation}\label{Eq:Complex}
{\cal C}_\zeta: 0\rightarrow {\cal A}^0\xrightarrow{\delta_0}{\cal A}^1\xrightarrow{\delta_1}{\cal A}^2\rightarrow 0,
\end{equation}
with 
\begin{align}
\delta_0:  f \mapsto
\left(\begin{array}{c}   
-(D_t+\zeta Z^\dagger)f\\
(-Z+\zeta D_t^\dagger) f\\
(J+\zeta I^\dagger)f_{\Lambda_0}\\
(B^{LR}-\zeta (B^{RL})^\dagger)f_R+(b^{RL}+\zeta (b^{LR})^\dagger)f_L\\
(b^{LR}-\zeta (b^{RL})^\dagger)f_R+(B^{RL}+\zeta (B^{LR})^\dagger)f_L
\end{array}\right),
\end{align}
and
\begin{multline}
\delta_1=(-Z+\zeta D_t^\dagger , D_t+\zeta Z^\dagger)
+\sum_{\lambda\in\Lambda_0}\delta(s-\lambda)(I_\lambda-\zeta J_\lambda^\dagger)\\
+\sum_{e\in{\cal E}}\Bigg(
\delta(s-t(e))(-B_e^{RL}-\zeta(B_e^{LR})^\dagger)
	+\delta(s-h(e))(b_e^{LR}-\zeta(b_e^{RL})^\dagger)\\
\shoveright{\hfill-\delta(s-t(e))(b_e^{RL}+\zeta (b_e^{LR})^\dagger)  
			+\delta(s-h(e))( B_e^{LR}+\zeta (B_e^{RL})^\dagger)\Bigg),}
\end{multline}
i.e. for $\psi\in\Gamma(S\otimes E\otimes e), \chi\in W, v_{-}\in E_L\otimes_{\cal E} e_R,$ and $v_{+}\in E_R\otimes_{\cal E} e_L$ we have
\begin{multline}
\delta_1:\left(\begin{array}{c} \psi\\ \chi \\ v_-\\ v_+   \end{array}\right)\mapsto 
(-Z+\zeta D_t^\dagger)\psi_1+ (D_t+\zeta Z^\dagger)\psi_2
+\sum_{\lambda\in\Lambda_0}\delta(s-\lambda)(I_\lambda-\zeta J_\lambda^\dagger)\chi_\lambda\\
+\sum_{e\in{\cal E}}\Bigg(
\delta(s-t(e))(-B_e^{RL}-\zeta(B_e^{LR})^\dagger)v^e_-
	+\delta(s-h(e))(b_e^{LR}-\zeta(b_e^{RL})^\dagger)v^e_-\\
\shoveright{\hfill-\delta(s-t(e))(b_e^{RL}+\zeta (b_e^{LR})^\dagger)v^e_+  
			+\delta(s-h(e))( B_e^{LR}+\zeta (B_e^{RL})^\dagger)v^e_+\Bigg),}
\end{multline}

The exactness condition  
$\delta_1\delta_0=0$ at ${\cal A}^1$  is satisfied for all $\zeta$ if and only if 
\begin{equation}
\mu_{\mathfrak L}(Q,B,T)\otimes 1_e+1_E\otimes\mu_{\mathfrak s}(b,t)=0,
\end{equation}
which is exactly our choice of the matching of the moment map values of Eq.~\eqref{Matching}.  

Basic Hodge theory becomes very powerful in this setup if one observes that Eq.~\eqref{nDd} implies that the Dirac operator is directly related to the operator $\delta_{1}-\delta_{0}^\dagger: {\cal A}^1\rightarrow {\cal A}^0\oplus{\cal A}^2,$ namely
\begin{equation}
\left(\begin{array}{cc}
-1 & -\bar{\zeta}\\
-\zeta & 1
\end{array}\right){\cal D}_{t}^\dagger=\delta_{1}-\delta_{0}^\dagger.
\end{equation}
This implies that ${\cal D}n^{-1}\sim\delta_1^\dagger-\delta_0.$  As we argued in Section~\ref{Sec:NahmTransform}, away from the special locus $D$ the equation ${\cal D}\Psi=0$ has no solutions\footnote{{\rm Ker D} is empty, since, as we argued, the operator $D^\dagger D$ is strictly positive away from the divisor $D$ and ${\rm Ker} D^\dagger D=0.$  }.  It follows that both ${\rm Ker}\, \delta_0$ and ${\rm Ker}\, \delta_1^\dagger$ are empty and thus $H^0({\cal C}_\zeta)=0$ and $H^2({\cal C}_\zeta)=0.$

Now, similarly to the case of instantons on a four-torus discussed in \cite{DK}, we argue that the only interesting cohomology of ${\cal C}_\zeta$ can be identified with the kernel of ${\cal D}_t^\dagger$ operator $H^1({\cal C}_\zeta)={\rm Ker}\, {\cal D}_t^\dagger.$  It is clear that any element of $\psi\in{\rm Ker}\, {\cal D}_t^\dagger$ is in ${\rm Ker}\,\delta_1,$ since, due to Eq.~\eqref{nDd} $\delta_1\psi=0$ is the second component of the equation $n{\cal D}_t^\dagger\psi=0.$

The opposite inclusion ${\rm Ker}\, \delta_1\subset{\rm Ker}\, {\cal D}^\dagger$ follows from the following argument.  For any representative $\eta\in{\rm Ker}\, \delta_1$ we consider $\eta+\delta_0\rho$ which is also in ${\rm Ker}\, \delta_1.$ We seek $\rho$ satisfying the following condition  $\delta_0^\dagger(\eta+\delta_0\rho)=0.$ Since ${\rm Ker}\, \delta_0=0$ the operator $\delta_0^\dagger\delta_0$ is invertible and we can solve for  
$\rho=-(\delta_0^\dagger\delta_0)^{-1}\delta_0^\dagger\eta.$

If one views the Nahm transform as a form of the Fourier transform, then  the fact that it is an isometry is a form  of the Plancherel theorem.  This fact has been proved for the Nahm transform of instantons on a four-torus in \cite{Braam:1988qk}.  For the original ADHM construction of instantons on ${\mathbb R}^4$ it was proved in \cite{Maciocia:1991ph}, for instantons on ALE spaces in \cite{KN}, for the monopoles in \cite{Nakajima:1990zx}, and for doubly-periodic instantons in  \cite{Biquard:2000ed}.  As we demonstrated above, the Nahm transform of Section \ref{Sec:NahmTransform} is a triholomorphic isomorphism.  To prove that it is an isometry it remains to be shown that the holomorphic two-form is also preserved.  Based on the cases listed and on the fact that in our case the Nahm transform still respects all of the complex structures, we work now under the hypothesis that the moduli space of instantons is isometric to the moduli space of the corresponding bow representation.  We now turn to study these spaces.

\section{Moduli Spaces of Instantons}
\label{Sec:HKR}
The moduli space of a bow representation was defined so far as an infinite hyperk\"ahler quotient, i.e. the space appeared as a hyperk\"ahler reduction of an infinite-dimensional affine space by an infinite-dimensional group.   There are various ways of writing each such moduli space as a finite   hyperk\"ahler quotient, each useful and interesting in its own right.

\subsection{Moduli Space of Instantons as a Finite Hyperk\"ahler Quotient}
Limiting our attention to some subinterval $I_\sigma^\alpha$  of length $l_\sigma^\alpha=\lambda_\sigma^{\alpha+1}-\lambda_\sigma^\alpha,$ in any given trivialization of $E_\sigma^\alpha\rightarrow I_\sigma^\alpha$  the space of the Nahm data associated with it consists  of the Nahm matrices $T_0, T_1, T_2, T_3$ of size $R_\sigma^\alpha\times R_\sigma^\alpha$ and boundary conditions \eqref{JumpMatch}, such that $T_0$ is regular and $T_j$ have poles at each end of the interval forming representations of dimension $R_\sigma^\alpha-R_\sigma^{\alpha+1}$ at the right end, if this quantity is positive, and of dimension   $R_\sigma^\alpha-R_\sigma^{\alpha-1},$ if this quantity is positive. If either of these rank changes is not positive, then all $T_i$ are regular at the corresponding end.  Also, in the case $\alpha=0$ or $r_\sigma$ all $T_j$ are regular at, respectively, the left or right end.  Let us denote the space of the Nahm data satisfying these conditions and solving the Nahm equations within the interval modulo the action of the group of the gauge transformations which are trivial at the ends by ${\cal O}_{R_\sigma^\alpha}(R_\sigma^{\alpha-1}, R_\sigma^{\alpha+1}; l_\sigma^\alpha).$  These hyperk\"ahler manifolds were extensively studied; see for example \cite{Bielawski:1997}, where they are described as submanifolds in $T^*Gl(n, {\mathbb C}).$   To mention one such space, in any given complex structure ${\cal O}_n(n,n,l)$ is $T^*Gl(n,{\mathbb C})=Gl(n,{\mathbb C})\times gl(n,{\mathbb C})$ with Kirillov-Konstant form as its complex symplectic form \cite{Kronheimer:1988}.  

Consider the group ${\cal G}_0$ of gauge transformations that have trivial action at the ends of all of the subintervals
\begin{equation}
{\cal G}_0=\left\{ g \Big| g(p_{\sigma R})=g(p_{\sigma L})=g(\lambda)=0 \right\}.
\end{equation}
Performing the hyperk\"ahler reduction by ${\cal G}_0$ we reduce ${\rm Dat}={\cal N}\oplus{\cal F}_{\rm in}\oplus{\cal F}_{\rm out}\oplus{\cal B}\oplus\bar{\cal B}$ to the finite-dimensional space.  Reduction on each interval produces one of the spaces 
\begin{align}
{\cal O}_\sigma^\alpha&\equiv 
{\cal O}_{R_\sigma^\alpha}(R_\sigma^{\alpha-1},R_\sigma^{\alpha+1}; l_\sigma^\alpha)
,\ {\rm for}\ \alpha=1,2,\ldots, r_\sigma\\
{\cal O}_\sigma^0&\equiv  
{\cal O}_{R_\sigma^0}(R_\sigma^{0},R_\sigma^{1}; l_\sigma^0)
,\\
{\cal O}_\sigma^{r_\sigma}&\equiv  
{\cal O}_{R_\sigma^{r_\sigma}}(R_\sigma^{r_\sigma-1},R_\sigma^{r_\sigma}; l_\sigma^{r_\sigma}).
\end{align}
The Nahm data is the only component of ${\rm Dat}$ that transforms under ${\cal G}_0$ and 
\begin{equation}
{\cal N}/\!\!/\!\!/{\cal G}_0=\mathop{\oplus}_{\sigma\in{\cal I}}\mathop{\oplus}_{\alpha=0}^{r_\sigma} {\cal O}_\sigma^\alpha.
\end{equation}

It is clear that the group ${\cal G}/{\cal G}_0$ is the finite-dimensional group of gauge transformations at the ends of the intervals and at the $\lambda$-points.  Thus the total moduli space of a bow representation $\mathfrak R$ is the finite hyperk\"ahler quotient
\begin{equation}
{\cal M}_{\mathfrak R}={\cal F}_{\rm in}\oplus{\cal F}_{\rm out}\oplus{\cal B}\oplus\bar{\cal B}\mathop{\oplus}_{\sigma,\alpha} {\cal O}_\sigma^\alpha\Big/\!\!\!\Big/\!\!\!\Big/ \prod U_\sigma^\alpha\times U_L\times U_R,
\end{equation}
where $U_L=\prod_\sigma U(R_\sigma^0), U_R=\prod_\sigma U(R_\sigma^{r_\sigma}),$ and $U_\sigma^\alpha=U\big(\min\{R_\sigma^\alpha, R_\sigma^{\alpha-1}\}\big)$ for $0<\alpha<r_\sigma.$ 

The structure of the spaces  ${\cal O}_\sigma^\alpha$ is quite interesting and they provide a useful realization to the moduli space of the bow representation. Nevertheless, this is not the realization we find most useful in what follows.

\subsection{A Different Finite Quotient Realization}
Let us now focus on the moduli space of  a representation of an $A$-type bow.  Our results will generalize easily from this to the general case. Following Bielawski \cite{Bielawski:1998hk} and \cite{Bielawski:1998hj} we use the following spaces: $F_n(m,c)={\cal O}_n(m,n;c).$ A space $F_n(m,c)$ is the space of gauge equivalence classes of the file $n$ Nahm data on an interval of length $c$ that are regular on one end and have a $(n-m)\times(n-m)$ block with a pole at the other end with residues forming an irreducible representation of $su(2).$

On the other hand the general ${\cal O}_n(m_1, m_2; l)$ space can be represented in terms of $F$-spaces as ${\cal O}_n(m_1, m_2; l)=F_n(m_1,c)\times F_n(m_2, l-c)/\!\!/\!\!/U(n)$ for any positive $c<l.$

Let us choose a point $pt_\sigma^\alpha$ in each interval  $I_\sigma^\alpha,$ so for any $\sigma$ and $\alpha=0,\ldots,r_\sigma$ we have $pt_\sigma^\alpha\in I_\sigma^\alpha.$  Each interval $I_\sigma^\alpha$ is divided into two parts by $pt_\sigma^\alpha.$  We denote the length of the left part by $c_\sigma^\alpha,$ so $0\leq c_\sigma^\alpha\leq l_\sigma^\alpha,$ and the length of the right part by $l_\sigma^\alpha-c_\sigma^\alpha.$ Now we can write the moduli space ${\cal M}_{\mathfrak R}$ as a finite quotient in a different way.  Let $F_{m, m'} (c, c')=F_n(m,c)\times F_n(m',c')/\!\!/\!\!/U(\min\{m,m'\})$ for $m\neq m'$ and let $F_{m,m} (c,c')=F_m(m,c)\times F_m(m,c')\times{\mathbb C}^{2m}/\!\!/\!\!/U(m).$  We define these spaces as in \cite{Bielawski:1998hj}.  We need one more ingredient associated to each edge of the bow. This is the space  
$G_{m,m'}(c,c')=F_m(m,c)\times F_{m'}(m', c')\times{\mathbb C}^{2m m'}/\!\!/\!\!/U(m)\times U(m').$

Let ${\cal G}^c$ be the subgroup of $\cal G$ consisting of the gauge transformations that are identity at the chosen points $pt_\sigma^\alpha$:
\begin{equation}
{\cal G}^c=\left\{ g | g(pt_\sigma^\alpha)=1\right\}.
\end{equation}
Then ${\cal G}/{\cal G}^c=\prod_{\sigma, \alpha} U(R_\sigma^\alpha).$

For each interval consider the space 
\begin{equation}
{\rm Int}_\sigma=\prod_{\alpha=1}^{r_\sigma} F_{R_\sigma^{\alpha-1},R_\sigma^\alpha}(l_\sigma^{\alpha-1}-c_\sigma^{\alpha-1}, c_\sigma^\alpha)\bigg/\!\!\!\!\!\bigg/\!\!\!\!\!\bigg/\prod_{\alpha=1}^{r_\sigma} U(R_\sigma^\alpha).
\end{equation}
Now
\begin{equation}
{\cal M}_{\mathfrak R}=\prod_\sigma {\rm Int}_\sigma \times \prod_e 
G_{R_{t(e)}^{r_{t(e)}}, R_{h(e)}^0}(l_{t(e)}^{r_{t(e)}}-c_{t(e)}^{r_{t(e)}}, c_{h(e)}^0)\bigg/\!\!\!\!\!\bigg/\!\!\!\!\!\bigg/\prod \left(U(R_{t(e)}^{r_{t(e)}})\times U(R_{h(e)}^0)\right).
\end{equation}
Of course, one can choose $pt_\sigma^0=p_{\sigma L}$ and all $pt_\sigma^{r_\sigma}=p_{\sigma R}$ so that $c_\sigma^0=0$ and $c_\sigma^{r_\sigma}=l_\sigma^{r_\sigma}$ for every $\sigma.$ In this case
\begin{equation}
G_{R_{t(e)}^{r_{t(e)}}, R_{h(e)}^0}(l_{t(e)}^{r_{t(e)}}-c_{t(e)}^{r_{t(e)}}, c_{h(e)}^0)={\mathbb C}^{2 R_{h(e)}^0 R_{t(e)}^{r_{t(e)}}},
\end{equation} 
and we quotient the product of $F_{m,m'}, F_m,$ and ${\mathbb C}^{2m m'}$ types of spaces only.  For any regular representation of a general bow, not necessarily of $A$-type, we have the following formula
\begin{equation}
{\cal M}_{\mathfrak R}=\prod_{\sigma\in{\cal I}}{\rm Int}_\sigma\times\prod_{e\in{\cal E}} {\mathbb C}^{2 R_{h(e)}^0 R_{t(e)}^{r_{t(e)}}}
\bigg/\!\!\!\!\!\bigg/\!\!\!\!\!\bigg/
\prod_{\sigma\in{\cal I}}\left(U(R_\sigma^0)\times U(R_\sigma^{r_\sigma})\right).
\end{equation}

To demonstrate the usefulness of this realization let us compute the dimension of the moduli spaces of a representation of an $A$-type bow, if this moduli space is not empty.

For $m<n$ the space $F_n(m, c)$ is biholomorphic to $Gl(n, {\mathbb C})\times gl(m, {\mathbb C})\times{\mathbb C}^{n+m}$ \cite{Hurtubise:1989wh, Bielawski:1997} and
\begin{equation}
\dim_{\mathbb C} F_n(m, c)=n(n+1)+m(m+1),
\end{equation}
while for $m=n$ the space $F_n(n, c)$ is biholomorphic to $Gl(n, {\mathbb C})\times gl(n, {\mathbb C})$ and 
\begin{equation}
\dim_{\mathbb C} F_n(n, c)=2n^2.
\end{equation}

For any $n$ and $m\leq n$
\begin{equation}
\dim_{\mathbb C} F_{m, m'}(c,c')=m(m+1)+m'(m'+1). 
\end{equation}
Since for $m=m'$ we have $F_{m, m}(c,c')=F_m(m,c)\times F_m(m,c')\times{\mathbb C}^{2m}/\!\!/\!\!/U(m),$ this gives $\dim_{\mathbb C} F_{m, m}(c,c')=2m^2+2m^2+2m-2m^2=2m(m+1).$ For the case $m<m'$ we have instead
$F_{m, m}(c,c')=F_m(m,c)\times F_m'(m,c')/\!\!/\!\!/U(m)$ which gives $\dim_{\mathbb C} F_{m, m'}(c,c')=2m^2+m(m+1)+m'(m'+1)-2m^2=m(m+1)+m'(m'+1).$  The case $m>m'$ is completely analogous leading to the same answer.

The dimensions of the spaces $G_{m,m'}(c,c')=F_m(m,c)\times F_{m'}(m', c')\times{\mathbb C}^{2m m'}/\!\!/\!\!/U(m)\times U(m')$ associated to the edges are $\dim_{\mathbb C} G_{m,m'}(c,c')=2m^2+2(m')^2+2m m'-2m^2-2(m')^2$ giving
\begin{equation}
\dim_{\mathbb C} G_{m,m'}(c,c')=2 m m'.
\end{equation}

Consider some interval $I_\sigma.$  In order to avoid cumbersome notation we choose to suppress the $\sigma$ subscript on all relevant quantities. Since the internal space associated to $I_\sigma$ is ${\rm Int}_\sigma=F_{R^0, R^1}(l^0-c^0, c^1)\times F_{R^1, R^2}(l^1-c^1, c^2)\times\ldots
F_{R^{r-1}, R^r}(l^{r-1}-c^{r-1}, c^r)/\!\!/\!\!/U(R^1)\times U(R^2)\times\ldots U(R^r)$ its dimension is 
\begin{align}
\dim_{\mathbb C} {\rm Int}&=R^0(R^0+1)+R^r(R^r+1)+\sum_{\alpha=1}^{r-1} \left(2R^\alpha(R^\alpha+1)-2(R^\alpha)^2\right)\\
&=R^0(R^0+1)+R^r(R^r+1)+\sum_{\alpha=1}^{r-1} 2 R^\alpha.
\end{align}

For an edge $e\in{\cal E}$ let $\Delta R_e=|R_{t(e)}-R_{h(e)}|.$  Assembling the internal spaces ${\rm Int}_\sigma$ with the spaces $G_{R_{t(e)}, R_{h(e)}}$ associated with the edges we find the dimension 
\begin{equation}
\dim_{\mathbb C} {\cal M}_{\mathfrak R}=2\sum_\sigma\sum_{\alpha=1}^{r_\sigma-1} R_\sigma^\alpha+2\sum_{e\in{\cal E}}\min\{R_{t(e)}^{r_{t(e)}}, R_{h(e)}^0\}-
\sum_{e\in{\cal E}}\Delta R_e(\Delta R_e-1).
\end{equation}

\section{Asymptotic Metric on the Moduli Space of Instantons on a multi-Taub-NUT Space}
\label{Sec:Asymp}
\subsection{Balanced Bifundamentals}\label{Sec:Balanced}
Let us first consider an $A_{k-1}$ bow representation, such that for each edge the rank of the fiber at its head equals to the rank of the fiber at its tail.  We call such a representation a {\em balanced representation.} 
All other moduli spaces can be reduced to these.  For the $A_{k-1}$ diagram this condition means that for any $\sigma=1, \ldots, k$ the ranks of $E_\sigma^{r_\sigma}$ and $E_{\sigma+1}^0$ are equal: $R_\sigma^{r_\sigma}=R_{\sigma+1}^0$ and $R_k^{r_k}=R_1^0.$  

Until now we were labeling the subintervals by pairs $(\sigma, \alpha)$ with $\sigma=1, \ldots, k$ and $\alpha=0,1, \ldots, r_\sigma.$  For our purposes here it is convenient to label the subintervals between the $\lambda$-points by $\tau=1, \ldots, n$ in the following manner.  Consider a map $(\sigma, \alpha)\mapsto \tau$ given by 
\begin{equation}
\tau=\sum_{\rho=1}^{\sigma-1} r_\sigma+\alpha.
\end{equation}
This map is invertible if $\alpha\neq 0$ and $\alpha\neq r_\sigma$ and it maps the pairs $(\sigma, r_\sigma)$ and $(\sigma+1, 0)$ to the same value of $\tau.$ When invertible, the inverse map is given by
\begin{align}
\sigma(\tau)&=\min\Big\{\rho\, \big|\, \tau\leq\sum_{\rho'=0}^\rho r_{\rho'}\Big\},&
\alpha(\tau)&=\tau-\sum_{\rho=0}^{\sigma-1} r_\rho.
\end{align}
Let $l_\tau$ be the length of the corresponding subinterval  $I_\sigma^\alpha$ if $\alpha\neq 0, r_\sigma$ and let it be the sum of the lengths of $I_\sigma^{r_\sigma}$ and $I_{\sigma+1}^0$ otherwise.  

Let us define integers $M_\sigma^\gamma=R_\sigma^\gamma $ with $\gamma =0,1,\ldots,r_\sigma-1.$  These integers are the ranks of the bundles on all subintervals except for the rightmost subintervals $I_\sigma^{r_\sigma}.$  Let $M_\sigma=\sum_{\gamma =0}^{r_\sigma-1} M_\sigma^ \gamma$ and $M=\sum M_\sigma.$ We also let $l_\sigma^\beta$ denote the length of the subinterval $I_\sigma^\beta$ for $\beta=1,2,\ldots, r_\sigma-1$ and $l_\sigma^{0}$ be the sum of the lengths of $I_{\sigma-1}^{r_{\sigma-1}}$ and $I_{\sigma}^0.$   The moduli space $\cal M$ of this bow representation has quaternionic dimension $M.$

We now construct a certain space ${\cal M}_{\rm ap}$ associated to this bow representation and an incomplete hyperk\"ahler metric $g_{\rm ap}$ on it that serves as an exponentially good approximation to the moduli space metric in some asymptotic regions.  We begin by assigning $M_\sigma^\gamma$ points in ${\mathbb R}^3$ to each subinterval with with $\gamma =0,\ldots, r_\sigma-1$ (the rightmost subintervals excluded).  This is exactly the number equal to the rank of the corresponding vector bundle.  Each point has coordinate vector $\vec{x}_j$ and an associated phase $\tau_j\sim\tau_j+4\pi.$ We shall treat points belonging to the same subinterval as undistinguishable, thus we'll have to divide by the corresponding direct product of symmetric groups $S=\prod_\sigma\prod_{\gamma=0}^{r_\sigma-1} S_{M_\sigma^\gamma}$ to obtain $M_{\rm ap}.$    

We have a total of $M$ points numbered by $j=1,\ldots, M$ and we introduce functions $f(j)$ and $t(j)$ such that the point $i$ is associated with the subinterval $I_\sigma^\gamma$ if and only if $\sigma=f(i)$ and $\gamma=t(i).$  Let us refer to $f(i)$ as the  flavor of a point $i$ and $t(i)$ as the type of a point $i.$ It is also convenient to introduce $g(i)=t(i)+\sum_{\sigma=0}^{f(i)-1}M_\sigma.$  These functions can be defined by 
\begin{align}
f(i)&=\min \left\{\rho\, \Big|\, i\leq\sum_{\sigma\leq\rho} M_\sigma\right\},\\
t(i)&=\min \left\{b\, \Big|\, i\leq\sum_{\sigma<f(i)}M_\sigma+\sum_{c\leq b} M_{f(i)}^c\right\},\\
g(i)&=\min \left\{b\, \Big|\, i\leq\sum_{c\leq b} M_{f(i)}^c\right\}.
\end{align}
Let $l(i)$ be the length of the associated interval or, if $t(i)=r_{f(i)},$ the sum of the lengths of the associated interval $I_{\sigma}^0$ and of $I_{\sigma-1}^{r_{\sigma-1}},$  i.e. $l(i)=l_{f(i)}^{t(i)}+\delta_{0,t(i)}l_{f(i)-1}^{r_{f(i)-1}}.$ 

We now introduce the hyperk\"ahler metric
\begin{align}\label{Asymp}
g_{\rm ap}=\Phi_{ij}d\vec{x}_i\cdot d\vec{x}^j+\left(\Phi^{-1}\right)_{ij}(d\tau_i+\omega_i)(d\tau_j+\omega_j),
\end{align}
with 
\begin{align}\label{Harmon}
\Phi_{ij}=
\begin{cases}
l(i)+\frac{\delta_{0,t(i)}}{|\vec{x}_i+\vec{\nu}_{f(i)}|}+\sum_{k\neq i}\frac{s_{ik}}{|\vec{x}_i-\vec{x}_k|}&\text{if}\ i=j,\\
-\frac{s_{ij}}{|\vec{x}_i-\vec{x}_j|}&\text{if}\ i\neq j,
\end{cases}
\end{align}
where $s_{ij}$ is given by the affine Cartan matrix $C_{\alpha\beta}$ of the instanton gauge group
\begin{equation}
s_{ij}=-C_{g(i),g(j)}=
\begin{cases}
-2&\text{if}\  f(i)=f(j)\, \text{and}\, t(i)=t(j),\\
1&\text{if}\  f(i)=f(j)\, \text{and}\, |t(i)-t(j)|=1,\\
1&\text{if}\ f(j)=f(i)+1, t(j)=0, t(i)=M_{f(i)-1},\\
1&\text{if}\ f(i)=f(j)+1, t(i)=0, t(i)=M_{f(j)-1},\\
0&\text{otherwise.}
\end{cases}
\end{equation}
We understand \eqref{Asymp} as a hyperk\"ahler metric on the space $\tilde{\cal M}$ of real dimension $4M$ fibered by $M$-dimensional tori $T^M$ over the configuration space of $M$ points in ${\cal R}^3.$ This metric clearly has $M$ triholomorphic isometries acting on the torus. The product $S$ of symmetric groups acts isometrically on $\tilde{\cal M}$ by permuting the points of the same type.  We define ${\cal M}_{\rm ap}$ as the quotient of $\tilde{\cal M}$ by this group: ${\cal M}_{\rm ap}\equiv\tilde{\cal M}/S.$

Let $D'\subset {\cal M}_{\rm ap}$ be the set of points with $\vec{x}_i=\vec{x}_j$ for any $i$ and $j$ with $g(i)=g(j).$ Then there is $D\subset{\cal M}$ and a bijection $\phi: {\cal M}\setminus D\rightarrow {\cal M}_{\rm ap}\setminus D'.$  Thus $\vec{x}_j, \tau_j$ up to the action of $S$ can serve as local coordinates on ${\cal M}\setminus D.$ Moreover, the metric $g$ on ${\cal M}$ is exponentially close to $g_{\rm app}$ of Eq.~\eqref{Asymp} in the following sense.
If $Rad=\min\{|\vec{x}_i-\vec{x}_j| \big| g(i)=g(j)\}$ is the minimal distance between the points of the same kind, then in the region of large $Rad$ the metric $g$ on $\cal M$ asymptotically approaches the metric $g_{\rm ap}$ exponentially fast, i.e. $|g-g_{\rm ap}|<\exp(-C Rad).$  

In order to prove this statement one can employ theorems and techniques developed by Bielawski in \cite{Bielawski:1998hk} and \cite{Bielawski:1998hj}.  This proof will appear in a forthcoming publication \cite{BC}.

\subsection{General Case}\label{Sec:GenRep}
Here we consider the case of general rank, i.e. we impose no relation on the ranks of $E_\sigma^{r_\sigma}$ and $E_{\sigma+1}^0.$  Let the difference of these two ranks be denoted by $\Delta R_\sigma=R_{\sigma+1}^0-R_\sigma^{r_\sigma},$  and let us introduce the quantities $c_\sigma^\tau$ by
\begin{equation}
c_\sigma^\tau=\sum_{\stackrel{\gamma\neq z(\sigma)}{\gamma=z(\sigma)}}^{z(\sigma)+\Delta R_\sigma}
|\Delta R_\sigma+z(\sigma)-\gamma|\delta((\gamma-\tau)\,  {\rm mod}\, n).
\end{equation}
Given the ranks 
\begin{equation}
R_\tau=\begin{cases}
R_{\sigma(\tau)}^{\beta(\tau)}&\text{if} \ \beta(\tau)\neq 0,\\
\min\Big\{R_{\sigma(\tau)}^{0}, R_{\sigma(\tau)-1}^{r_{\sigma(\tau)-1}}  \Big\}& \text{if}\ \beta(\tau)=0,
\end{cases}
\end{equation}
we let 
\begin{equation}
M_\tau=R_\tau-\sum_\sigma c_\sigma^\tau.
\end{equation}

If any $M_\tau$ is negative then the moduli space of this representation is empty.  If all $M_\tau$ are nonnegative then the moduli spaces has real dimension $4\sum_{\tau=1}^n M_\tau$ and the metric on it is given by 
Eqs.~\eqref{Asymp} and \eqref{Harmon}.

\section{Conclusions}
For every bow representation we define bow solutions.  The main result of this work is the formulation of the ADHM-Nahm transform that for a pair of bow representations ${\mathfrak L}$ and ${\mathfrak s}$  and a  given solution of ${\mathfrak L}$ constructs a bundle with a connection on the moduli space ${\cal M}_{\mathfrak s}$ of ${\mathfrak s}.$  We prove that in any of the complex structures on ${\cal M}_{\mathfrak s}$ the corresponding holomorphic bundle is flat, thus the curvature of this  connection is of type  $(1,1)$ in any of the complex structures.  Whenever ${\cal M}_{\mathfrak s}$ is four-dimensional this implies that the curvature is self-dual\footnote{This curvature is anti-self-dual in the conventional orientation chosen in the mathematics literature. See the footnote on page \pageref{Page:SDorASD}}.

Our main goal is to construct instantons on ALF spaces.  We realize each of these spaces as a moduli space ${\cal M}_{\mathfrak s}$ of a concrete bow representation.  Thus, for any solution of any other representation ${\mathfrak L}$ of the same bow, using the ADHM-Nahm transform we have formulated, one obtains an instanton solution on the ALF space ${\cal M}_{\mathfrak s}$.

The relation of the quiver moduli spaces to the representation theory is most intriguing.  This relation remains to be explored for the bow moduli spaces. In particular, our realization of the moduli space of a bow as a finite hyperk\"ahler quotient leads to a natural generalization of the notion of a quiver representation.  Certain bow representations, namely those that are reciprocal to the the balanced ones, produce all quiver representations in the limit of zero interval size. In this case a quiver has some vector spaces assigned to its vertices.  By exploring the limit of a general bow  representation one would obtain a new notion for a quiver representation. Now, the objects assigned to a quiver vertex would be not only vector spaces.  We expect these objects to be flag manifolds.

We did not treat a number of important analytic questions in this manuscript.  Among these are the proofs of finiteness of the action of the self-dual connection provided by the bow construction,  the analysis of the smoothness of the resulting connection, the isometry between the moduli space of self-dual connections and the moduli space of the large bow representation.  We imagine that the analytic techniques developed for treating similar questions in the case of calorons and instantons on ALE spaces extend to the current setup, however, it requires careful further analysis. 

If one is motivated to study the moduli spaces of instantons or the bow varieties by their relations to the representation theory or by the quantum gauge theories, then the question of $L^2$ cohomology of these spaces becomes important.  For a general study of instantons on ALF spaces  the compactification of  Hausel-Hunsicker-Mazzeo \cite{Hausel:2002xg} of the base space proved to be very useful.  We expect it to play a role in the study of the $L^2$ cohomology of the moduli spaces of instantons on these spaces as well.

We have defined the notions of a bow, its representation, and the moduli space of a representation.  There is an obvious limit of a bow in which it degenerates into a quiver.  In this limit  a representation for which the bundle rank does not change at any of the $\lambda$-points produces a representation of the corresponding quiver.  For such representations in any given complex structure the moduli space of the bow and the moduli space of the corresponding quiver are isomorphic as complex varieties.  We emphasize, however, that as differential manifolds, they are different.  In particular the $L^{2}$ cohomology of the bow representation is typically  larger than that of the corresponding quiver.  Other bow representations have interesting moduli spaces which do not degenerate to the conventional quiver varieties in the quiver limit.

\section*{Acknowledgments}
It is our pleasure to thank Edward Witten and Hiraku Nakajima for useful conversations and Juan Maldacena for his hospitality during our visits to the IAS, Princeton. We are grateful to the Institut des Hautes \'Etudes Scientifiques, Bures-sur-Yvette for hospitality during the completion of this work.  This work was supported in part by Science Foundation Ireland Grant No. 06/RFP/MAT050.  

\bibliographystyle{unstr}

\end{document}